\documentstyle[epsf]{mn}
\newcommand{\noi}{\noindent}

\title[Integrable models of galactic discs]
{Integrable models of galactic discs with double nuclei}
\author[M. A. Jalali and A. R. Rafiee]
        {M. A. Jalali \thanks{E-mail: jalali@iasbs.ac.ir} 
        \& A. R. Rafiee \\
        Institute for Advanced Studies in Basic Sciences, 
        P.O. Box 45195-159, Gava Zang, Zanjan, IRAN}

\begin{document}
\label{firstpage}
\maketitle

\begin{abstract} 
We introduce a new class of 2-D mass models, whose potentials
are of St\"ackel form in elliptic coordinates. Our model 
galaxies have two separate strong cusps that form double 
nuclei. The potential and surface density distributions 
are locally axisymmetric near the nuclei and become 
{\it highly} non-axisymmetric outside the nucleus. The 
surface density diverges toward the cuspy nuclei with the 
law $\Sigma \propto r^{-2}$. Our model is sustained by
four general types of regular orbits: {\it butterfly},
{\it nucleuphilic banana}, {\it horseshoe} and 
{\it aligned loop} orbits. Horseshoes and nucleuphilic bananas 
support the existence of cuspy regions. Butterflies 
and aligned loops control the non-axisymmetric shape of outer 
regions. Without any need for central black holes, our 
distributed mass models resemble the nuclei of M31 and 
NGC4486B. It is also shown that the self-gravity of the 
stellar disc can prevent the double nucleus to collapse.
\end{abstract}


\section{INTRODUCTION}
{\it Hubble Space Telescope} (HST) data revealed that M31 and NGC4486B 
have double nuclei (Lauer et al. 1996, hereafter L96; Tremaine 1995,
hereafter T95). M31 has a bright nucleus (P1) displaced from the 
centre of the isophotal lines of outer regions and a fainter nucleus 
(P2) just at the centre. NGC4486B exhibits a similar structure with
a minor difference: The centre of outer isophotes falls between
P1 and P2. There are some explanations for the   
emergence of the double nuclei of these galaxies, among which the 
eccentric disc model of T95 has been more impressive. In the 
model of T95, a central black hole (BH) enforces stars to move
on ``aligned" Keplerian orbits, which may elongate in the same
direction as the long-axis of the model. Stars moving on
aligned Keplerian orbits linger near apoapsis and may result 
in P1. The mass of central ``supermassive" BH should be much 
greater than the mass of neighboring disc. Otherwise, the 
asymmetric growth of P1 won't allow the BH to remain
in equilibrium. 

Goodman \& Binney (1984) showed that central massive
objects enforce the orbital structure of stellar systems 
to evolve towards a steady symmetric state. This result 
was then confirmed by the findings of Merritt \& Quinlan 
(1998) and Jalali (1999, hereafter J99) in their study of 
elliptical galaxies with massive nuclear BHs. 
Within the BH sphere of influence, highly non-axisymmetric 
structure can only exist for a narrow range of BH mass 
(J99). The results of J99 show that {\it long-axis tube} 
orbits of non-axisymmetric discs with central massive BHs,
elongate in the both $\pm$ directions of long-axis. 
Thus, the probabilities for the occurrence of two bright
regions, in both sides of BH along the long-axis, are
equal (these bright regions are supposed to be formed
near the apogee of long-axis tubes). By this hypothesis
one can interpret the double structure of NGC4486B by
placing a supermassive BH between P1 and P2. However,
some disadvantages arise in the case of M31. In 
the nucleus of M31, the formation of P1 can still be 
deduced from the behavior of long-axis tubes. But, 
there is no mathematical proof for the ``coexistence" of 
P1 and P2 when the centre of P2 coincides with BH's 
location.

In this paper we attempt to create a model based on the
self-gravity of stellar discs to show that systems with 
double nuclei can exist even in the absence of central BHs.   
Especial cases of our non-scale-free planar models are 
eccentric discs, which display a collection of properties 
expected in self-consistent cuspy systems.
Our models are of St\"ackel form in elliptic coordinates
(e.g., Binney \& Tremaine 1987) for which the Hamilton-Jacobi 
equation separates and stellar orbits are regular. 

In most galaxies, density diverges toward the centre in a 
power-law cusp. In the presence of a cusp, regular box orbits 
are destroyed and replaced by chaotic orbits (Gerhard \& Binney 1985). 
Through a fast mixing phenomenon, stochastic orbits cause the orbital 
structure to become axisymmetric at least near the centre
(Merritt \& Valluri 1996). These results are confirmed by the findings
of Zhao et al. (1999, hereafter Z99). Their study reveals that highly 
non-axisymmetric, scale-free mass models can not be constructed 
self-consistently. Near the cuspy nuclei, the potential functions
of our distributed mass models are proportional to $r^{-1}$ as
$r \rightarrow 0$. So, we attain an axisymmetric structure near
the nuclei which is consistent with the mentioned nature of
density cusps. The slope of potential function changes sign as
we depart from the centre and our model galaxies considerably 
become non-axisymmetric. Non-axisymmetric structure is supported
by {\it butterfly} and {\it aligned loop} orbits. Close to the 
larger nucleus, loop orbits break down and give birth to a new 
family of orbits, {\it horseshoe} orbits, which in turn generate 
{\it nucleuphilic banana} orbits. Stars moving in horseshoe and 
banana orbits lose their kinetic energy as they approach to the 
nuclei and contribute a large amount of mass to form cuspy 
regions. 

\section{THE MODEL}
Let us introduce the following family of planar potentials
expressed in the usual $(x,y)$ cartesian coordinates:
\begin{equation}
\Phi = K \frac {(r_1+r_2)^{\gamma}-
\beta(r_1-r_2)|r_1-r_2|^{\gamma-1}}{2r_1r_2}, \label{1}
\end{equation}
\begin{equation}
r_1^2 = (x+a)^2+y^2,~~r_2^2 = (x-a)^2+y^2, \label{2}
\end{equation}
\noi where $a$, $K>0$, $0 \le \beta \le 1$ and $2 < \gamma < 3$ 
are constant parameters. The points $(x=-a,y=0)$ and $(x=a,y=0)$
are the nuclei of our 2-D model. We call them P1 and P2, 
respectively. The distance between P1 and P2 is equal to $2a$.
The surface density distribution corresponding to $\Phi$
is determined by (Binney \& Tremaine 1987)
\begin{equation}
\Sigma (x',y')=\frac {1}{4 \pi ^2 G}
\int \int \frac {(\nabla ^2 \Phi) {\rm d}x{\rm d}y}
{\sqrt {(x'-x)^2+(y'-y)^2}}. \label{3}
\end{equation}
It is a difficult task to evaluate (\ref{3}) analytically.
So, we have adopted a numerical technique to calculate
this double integral. The functions $\Phi$ and $\Sigma$ 
are cuspy at P1 and P2. To verify this, we investigate the 
behavior of $\Phi$ and $\Sigma$ near the nuclei 
($r_1 \rightarrow 0$ and $r_2 \rightarrow 0$).
Sufficiently close to P1, we have $r_1 \ll r_2$ that 
simplifies (\ref{1}) as follows
\begin{equation}
\Phi=\frac {Kr_2^{\gamma -1}}2 
\frac {(1+\frac {r_1}{r_2})^{\gamma}+
\beta (1-\frac {r_1}{r_2})^{\gamma}}{r_1}. \label{4} 
\end{equation}
\noi We expand $(1+\frac {r_1}{r_2})^\gamma$ 
and $(1-\frac {r_1}{r_2})^\gamma$ in terms of $r_1/r_2$ to 
obtain
\begin{equation}
\Phi=\frac {Kr_2^{\gamma -1}}{2r_1} \sum_{n=0}^{\infty} 
\left [  
\frac {\left ( 1 + (-1)^n \beta \right ) \Gamma (\gamma+1)}
{n!\Gamma (\gamma -n+1)} \left ( \frac {r_1}{r_2} \right )^n 
\right ], \label{5} 
\end{equation}
\noi where $\Gamma$ is the well known Gamma function. 
As $r_1$ tends to zero, $r_2$ is approximated by $2a$ and
$r_1/r_2 \rightarrow 0$. Therefore, Equation (\ref{5}) reads
\begin{equation}
\Phi \approx \frac {Q(1+\beta)}{r_1},~~
Q=\frac 12 K(2a)^{\gamma-1}, \label{6}
\end{equation}
from which one concludes
\begin{equation}
\Sigma \propto Q(1+\beta)r_1^{-2}. \label{7}
\end{equation}
Similarly, it can readily be shown that the following 
approximations hold close to P2 ($r_2/r_1 \rightarrow 0$),
\begin{eqnarray}
\Phi &\approx& \frac {Q(1-\beta)}{r_2}, \label{8} \\
\Sigma &\propto& Q(1-\beta)r_2^{-2}. \label{9}
\end{eqnarray}
\noi In distant regions, when $r \gg a$ (with $r^2=x^2+y^2$), 
the potential function is approximated by
\begin{equation}
\Phi \approx 2^{\gamma-1}K r^{\gamma-2}. \label{10}
\end{equation}
\noi Correspondingly, 
\begin{equation}
\Sigma \propto r^{\gamma-3}. \label{11}
\end{equation}
\noi This shows that the surface density falls off outward if
$\gamma<3$. Besides, orbits will be bounded if the potential
$\Phi$ is concave in outer regions. This requirement
implies $\gamma >2$. Thus, we are restricted to $2 < \gamma < 3$. 
In Fig.~1, we have plotted the isocontours of 
$\Phi$ and $\Sigma$ for $\gamma=2.8$, $a=0.5$ and several
choices of $\beta$. The 3-D views of $\Phi$ and $\Sigma$ have 
also been demonstrated in Fig.~2. In \S~5, the potential surface 
of Fig.~2a will be referred as {\it potential hill}.

\begin{figure*}
\centerline{\hbox{\epsfxsize=2.0in\epsfbox{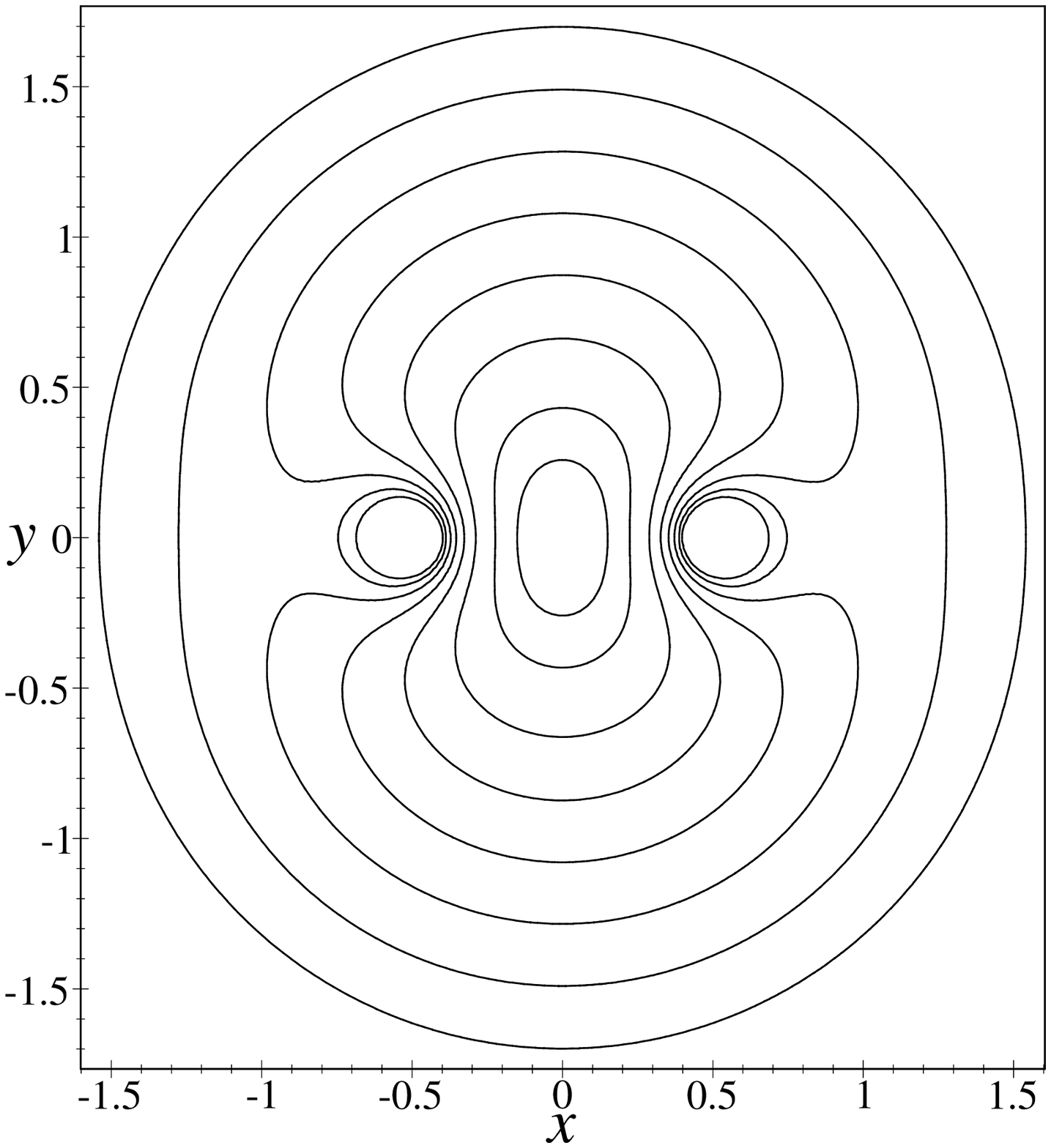}\hfill
                  \epsfxsize=2.0in\epsfbox{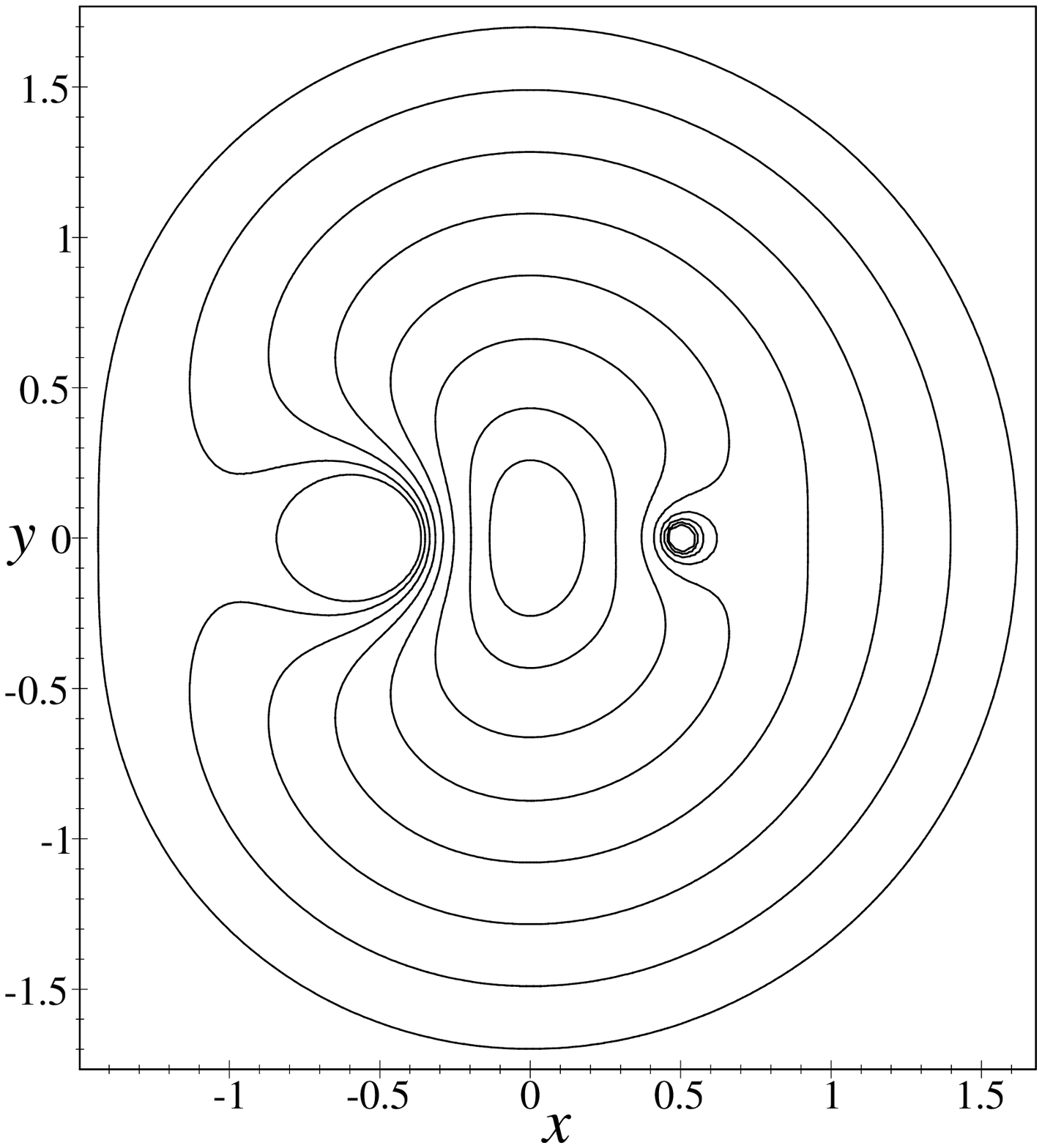}\hfill
                  \epsfxsize=2.0in\epsfbox{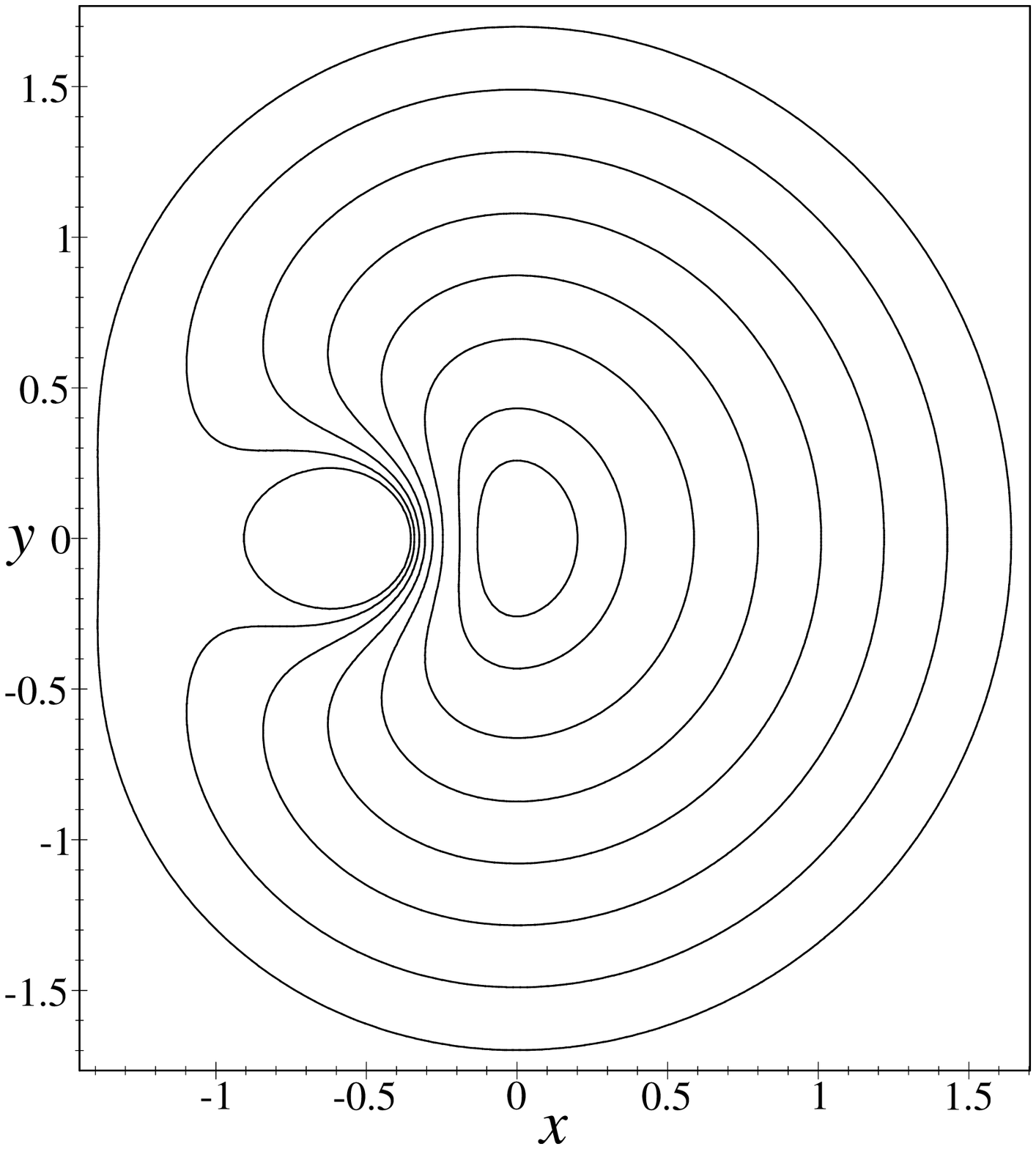}}}
\centerline{\hspace*{1.4in}$(a)$\hfill$(b)$\hfill$(c)$\hspace{1.4in}}
\centerline{\hbox{\epsfxsize=2.0in\epsfbox{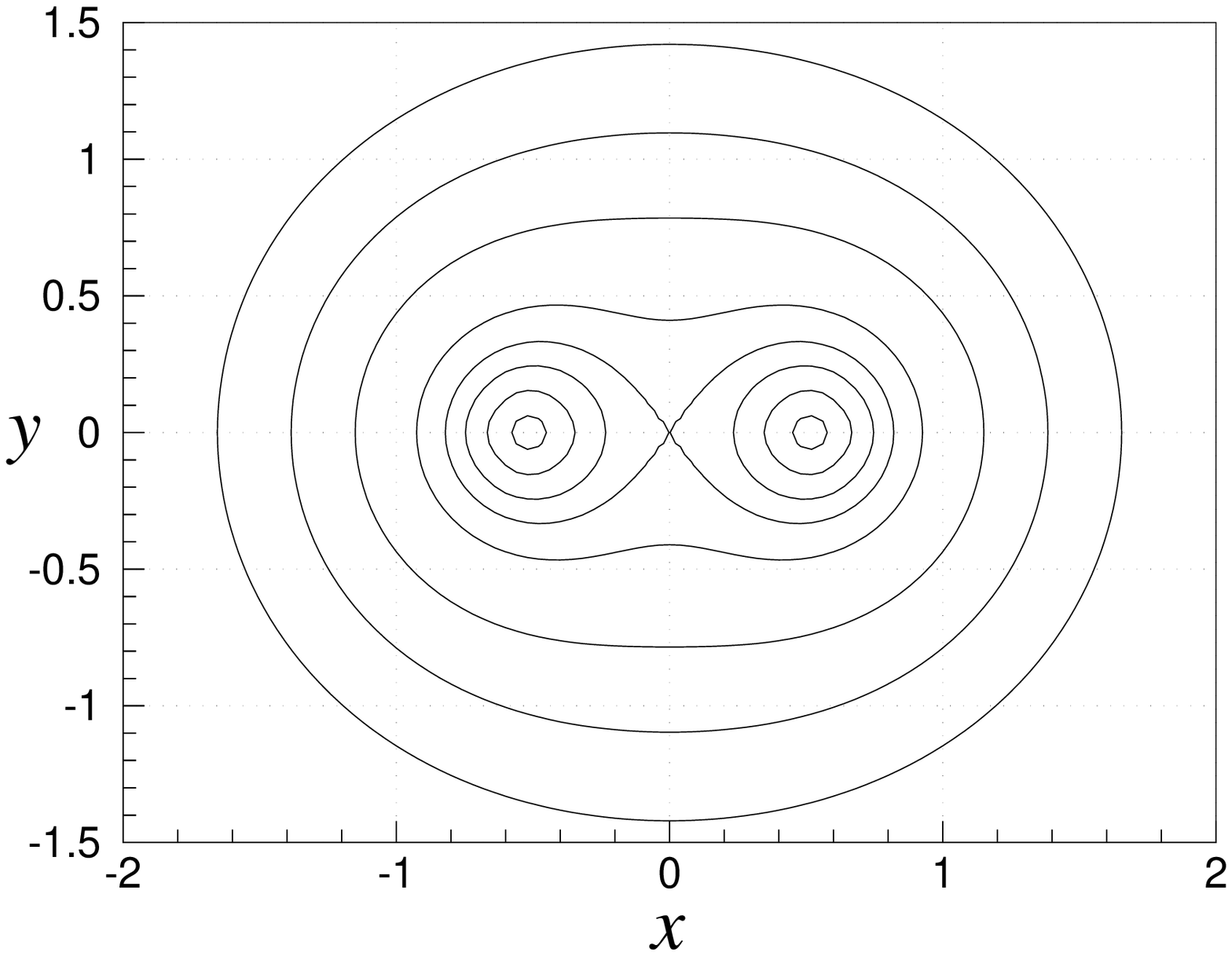}\hfill
                  \epsfxsize=2.0in\epsfbox{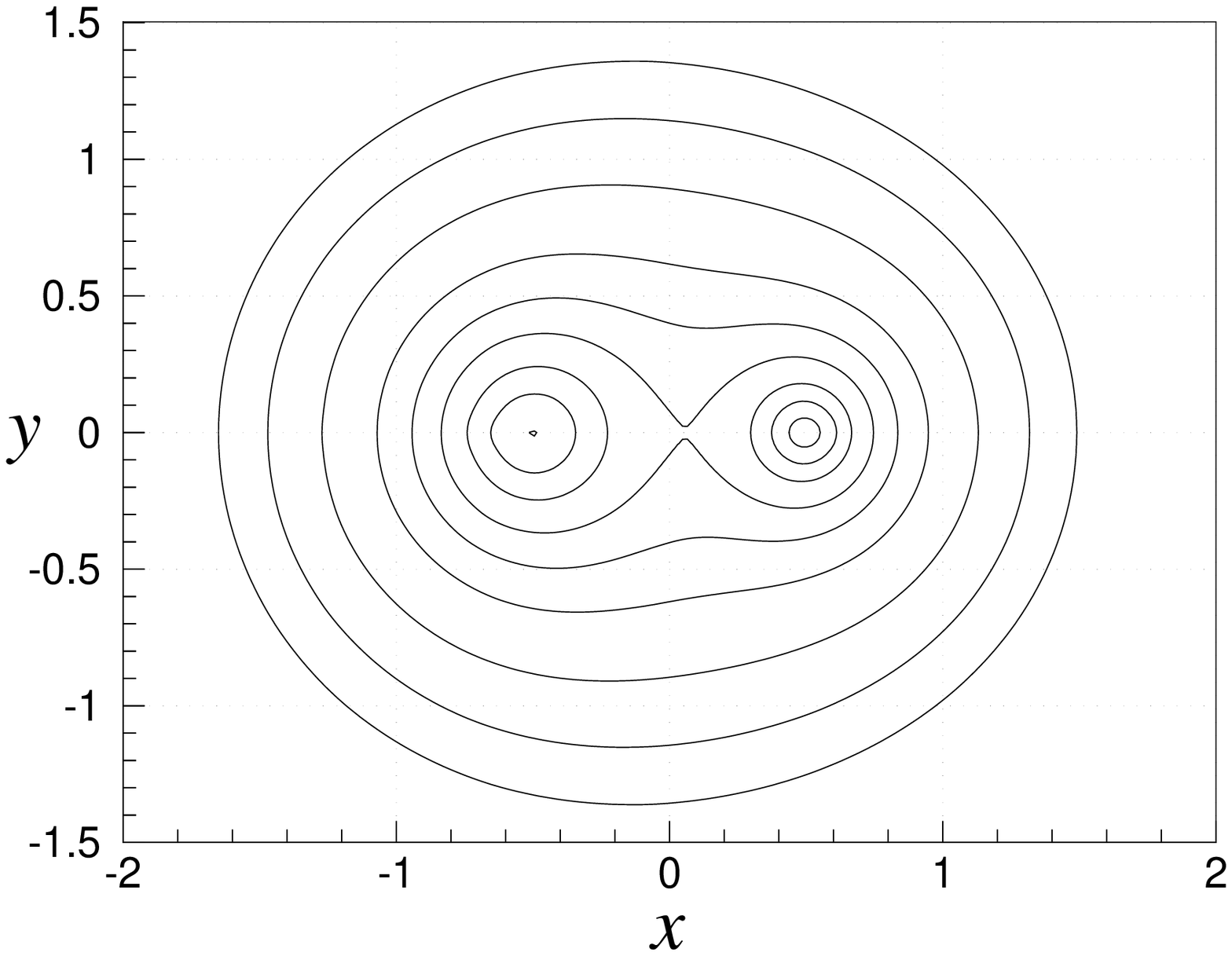}\hfill
                  \epsfxsize=2.0in\epsfbox{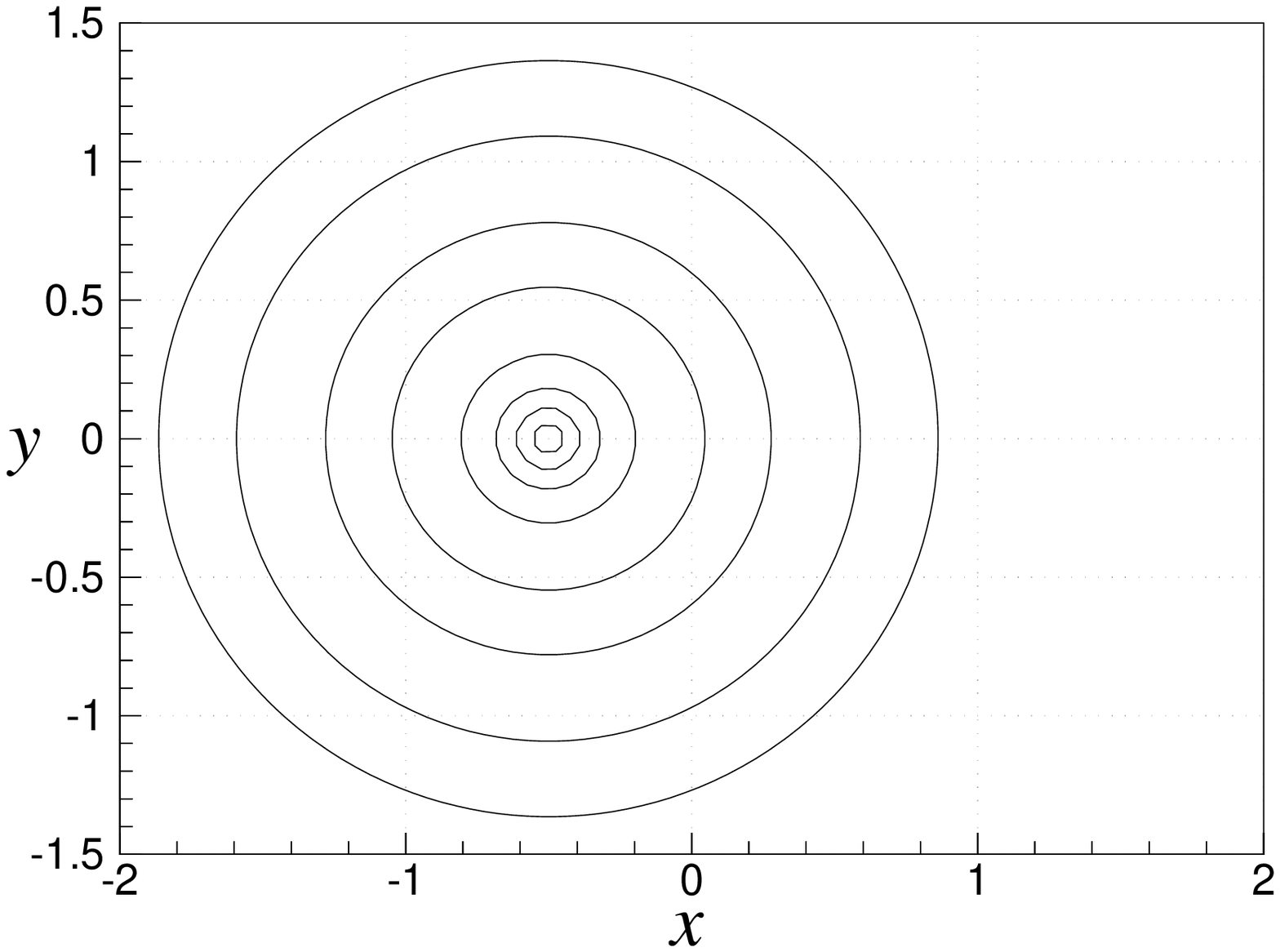}}}
\centerline{\hspace*{1.4in}$(d)$\hfill$(e)$\hfill$(f)$\hspace{1.4in}}
\caption[Figure 1]{The potential ($\Phi$) and surface density
($\Sigma$) isocontours for $\gamma=2.8$, $a=0.5$ and $K=0.2$. 
Figures (a), (b) and (c) show the potential isocontours 
for $\beta=0$, $\beta=0.75$ and $\beta=1$, respectively.
The corresponding surface density isocontours are
illustrated in Figures (d),(e) and (f).}
\end{figure*}

Figs.~1 and 2 assure that the potential 
and surface density functions are cuspy at P1 and P2. 
Regardless of the values of constant parameters,
the potential $\Phi$ has a local minimum at $(x=0,y=0)$. 
This minimum point can easily be distinguished in Figs.~1a, 1b 
and 1c. As the surface density isocontours show, the cuspy 
zones are disjointed by two separatrices that transversally 
intersect each other at a saddle point located on the 
$x$-axis between P1 and P2. The $x$-coordinate of this 
point can be determined through solving
\begin{equation}
\frac {\partial \Sigma (x,y)}{\partial x}=0,~~y=0, \label{12}
\end{equation}
\noi for $x$.  

The parameter $\beta$ controls the sizes of cuspy
zones around P1 and P2. For $\beta=0$, the sizes
of two cuspy zones are equal and the model has
reflection symmetries with respect to coordinate
axes. For $0<\beta <1$, the size of cuspy zone near
P1 is larger than that of P2 and the model is only
symmetric with respect to the $x$-axis.
The cuspy region around P2 is shrunk to zero size
when $\beta=1$ and we attain an eccentric disc with
a single nuclear cusp. 
Equations (\ref{6}) through (\ref{9}) show that
the potential and surface density functions are approximately
axisymmetric in the neighborhood of P1 and P2.
As we move outward, a ``highly" non-axisymmetric structure occurs.
For the large values of $r$, the surface density monotonically
decreases outward and our model galaxies become rounder again. 
Our mass models are indeed hybrid ones, which reflect the 
properties of density cusps and non-axisymmetric
systems, simultaneously. The centre of outer surface density 
isocontours falls at the middle of the centerline of P1 and P2. 
Nevertheless, the effective cuspy zones around P1 and P2 have 
different sizes.

\begin{figure*}
\centerline{\epsfxsize=3.0in\epsfbox{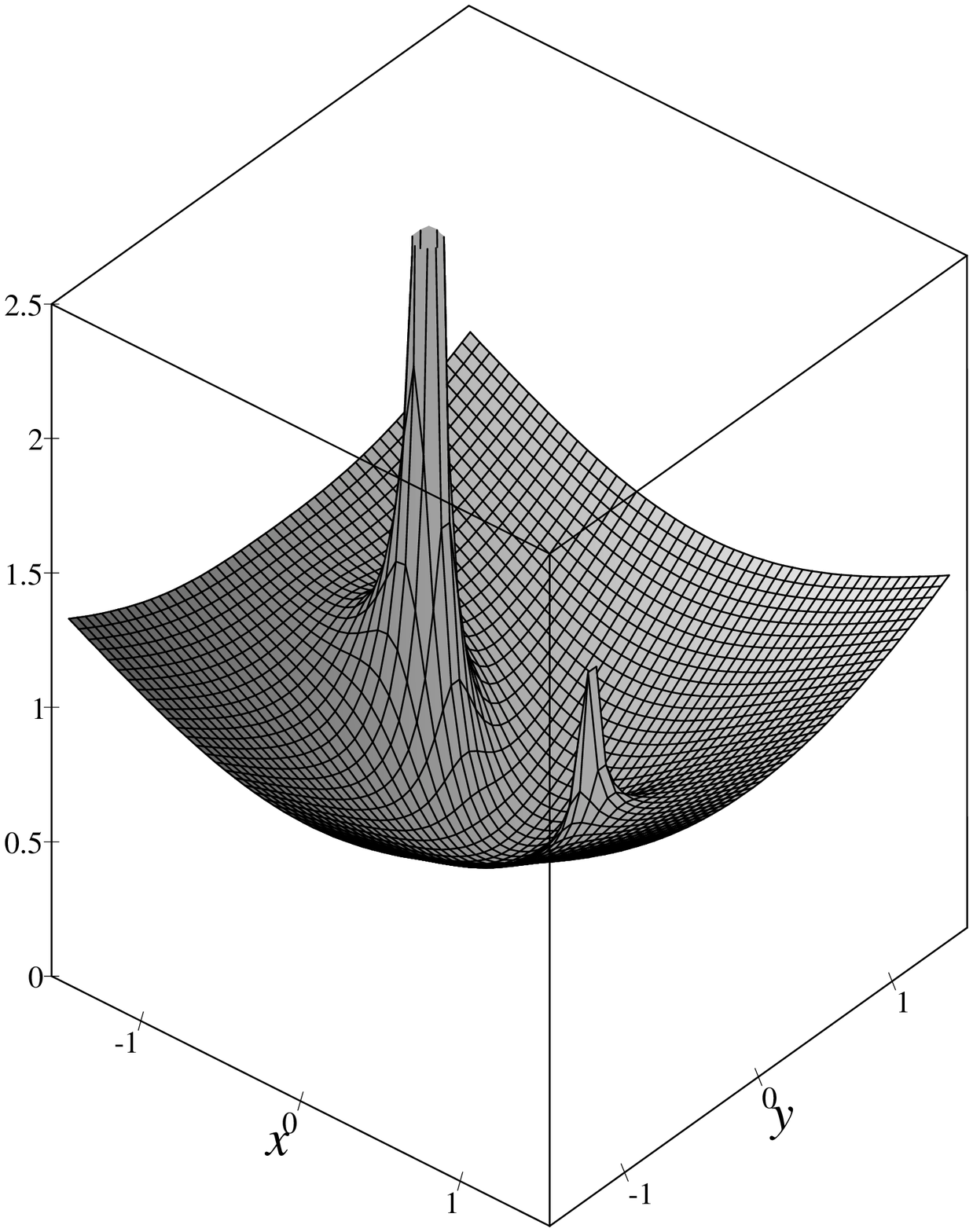}\hfill
            \epsfxsize=3.0in\epsfbox{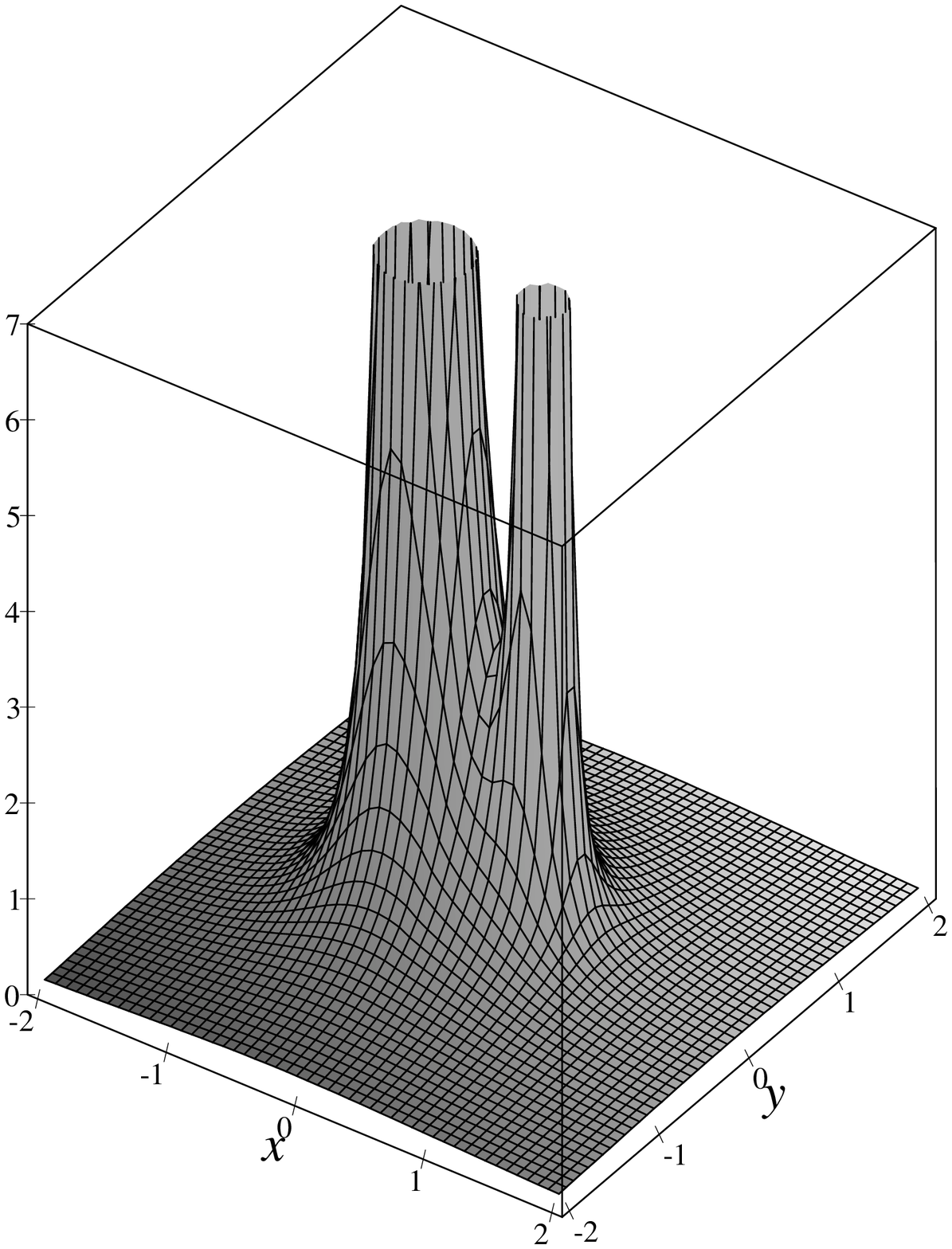}}
\centerline{\hspace*{1.5in}$(a)$\hfill$(b)$\hspace*{1.4in}}
\caption[Figure 2]{The 3-D views of $\Phi$ and $\Sigma$
for $\gamma=2.8$, $a=0.5$, $K=0.2$ and $\beta=0.75$. 
(a) the potential function (b) the surface density
distribution.} 
\end{figure*}

In what follows, we show that the potential $\Phi$ 
is of St\"ackel form in elliptic coordinates. We then 
classify possible orbit families, all of which are 
non-chaotic.

\begin{figure*}
\centerline{\epsfxsize=1.7in\epsfbox{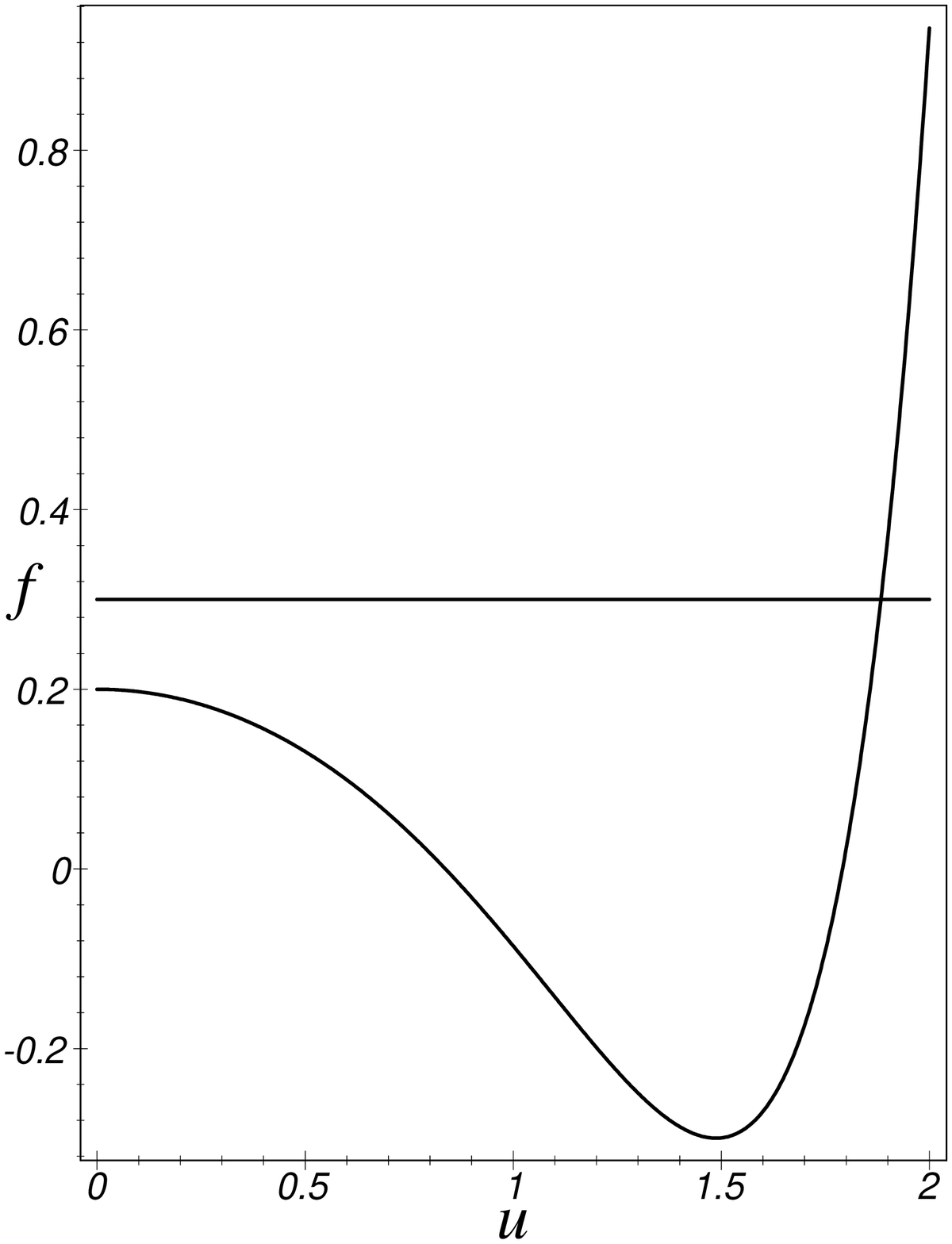}
            \epsfxsize=1.7in\epsfbox{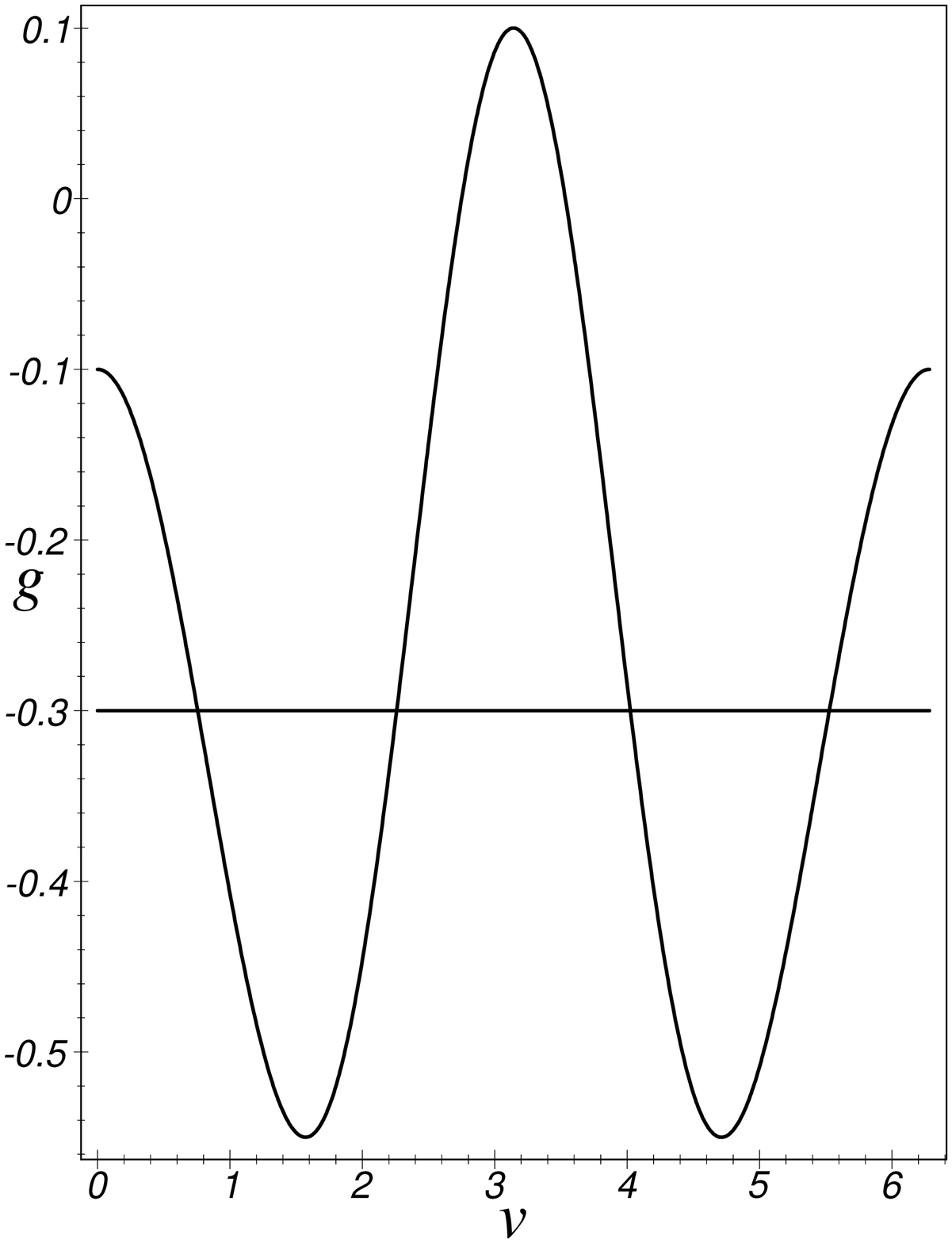}\hfill
            \epsfxsize=1.7in\epsfbox{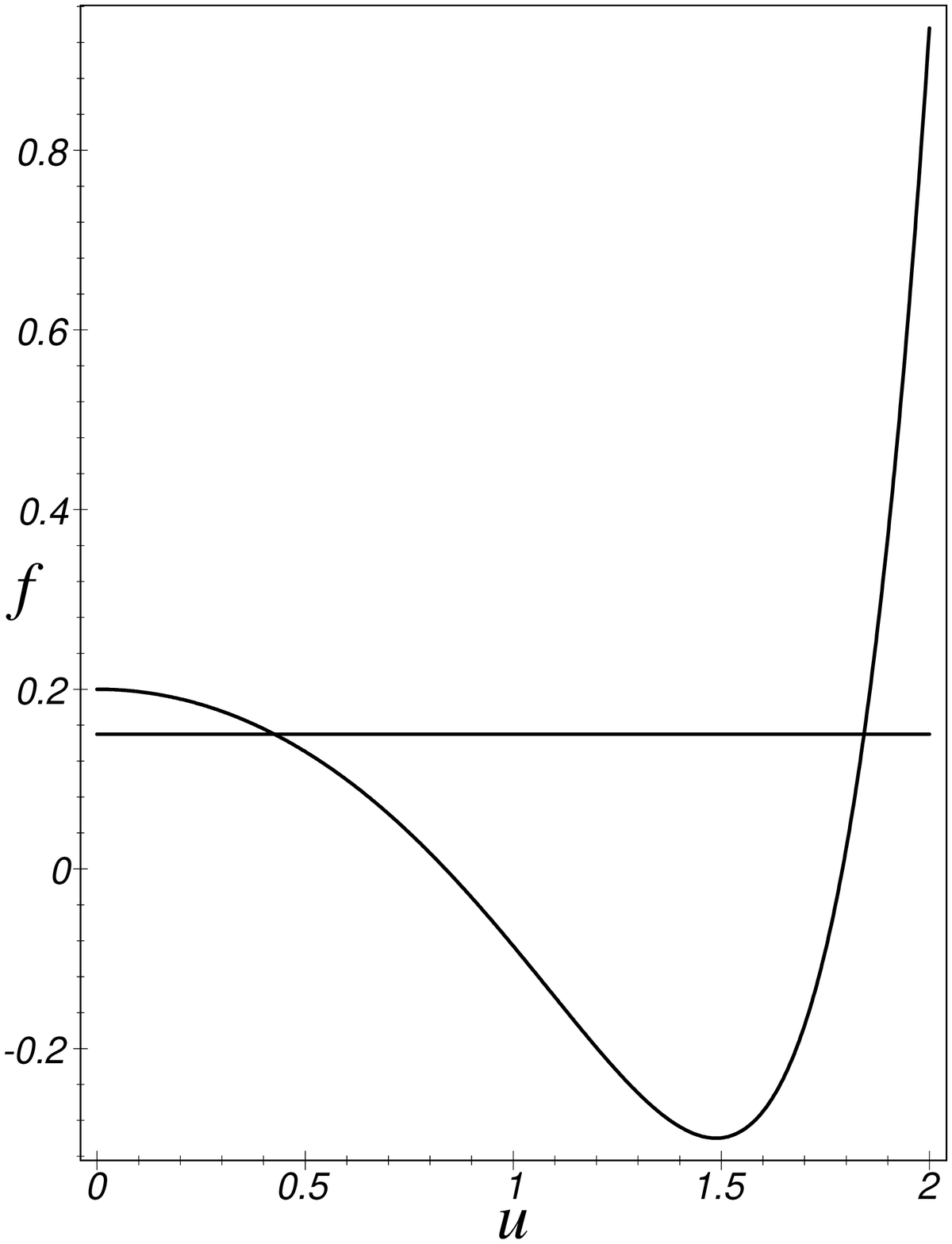}
            \epsfxsize=1.7in\epsfbox{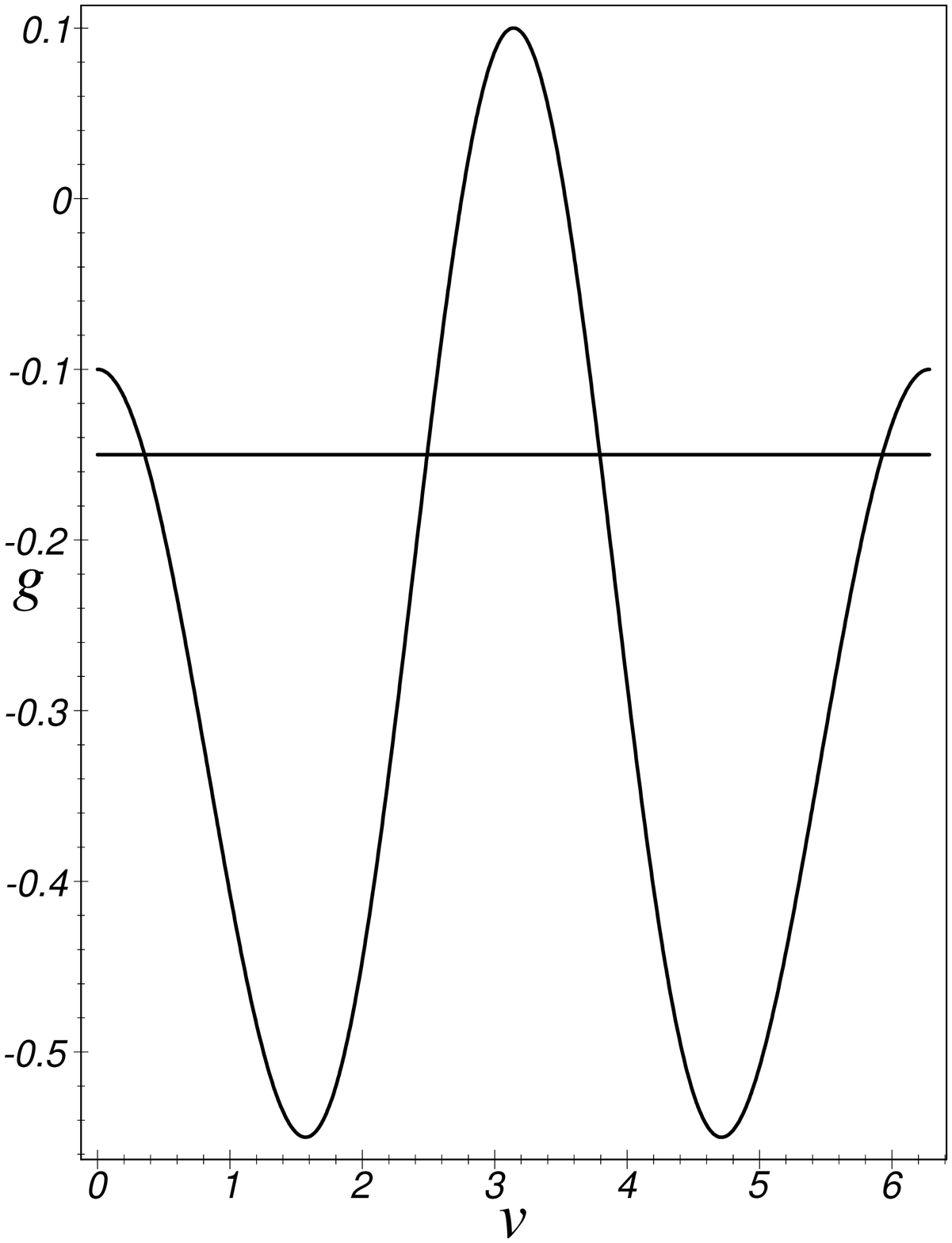}}
\centerline{\hspace*{0.8in}$(a)$\hspace*{1.5in}$(b)$\hfill
            $(c)$\hspace*{1.5in}$(d)$\hspace*{0.8in}}
\centerline{\epsfxsize=1.7in\epsfbox{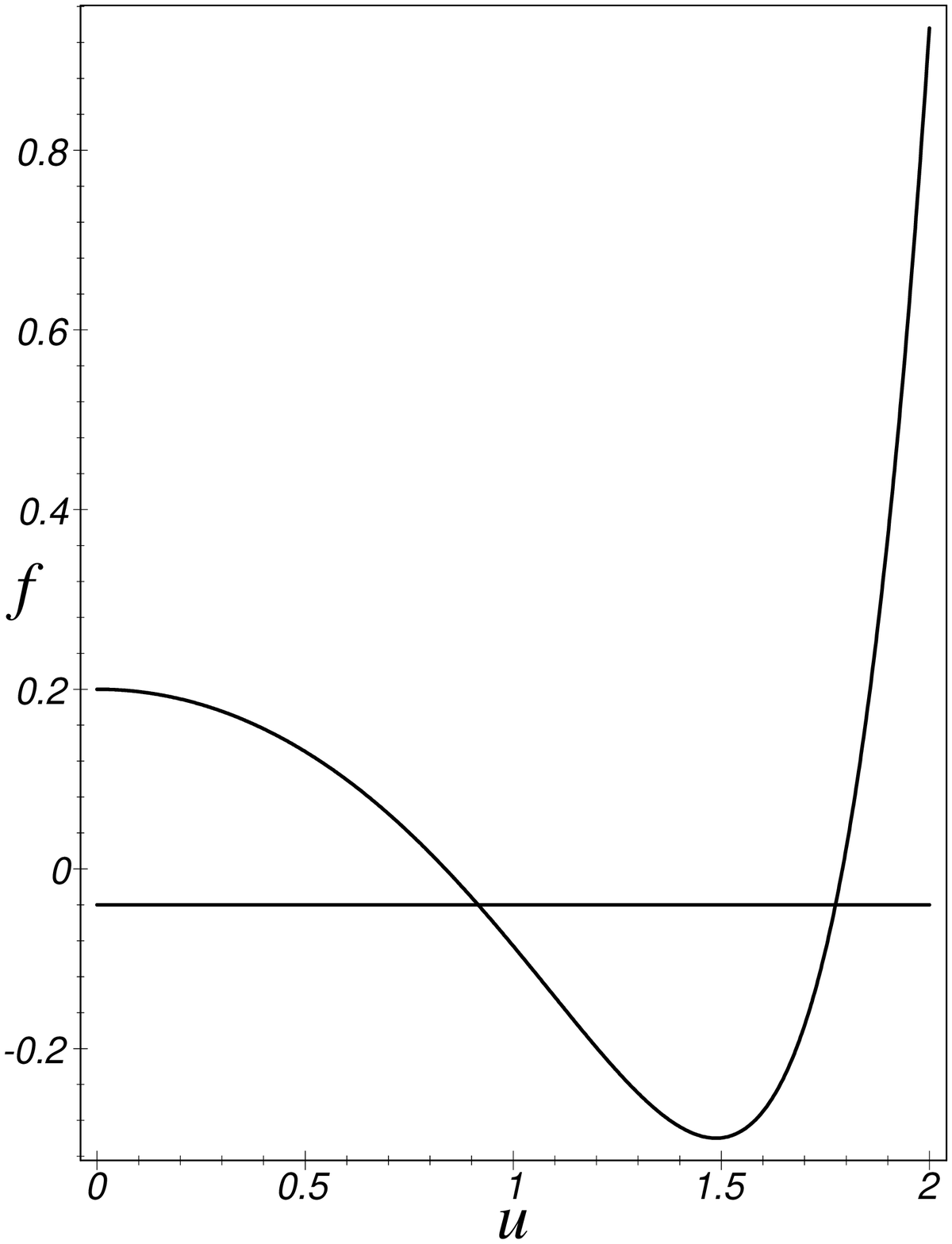}
            \epsfxsize=1.7in\epsfbox{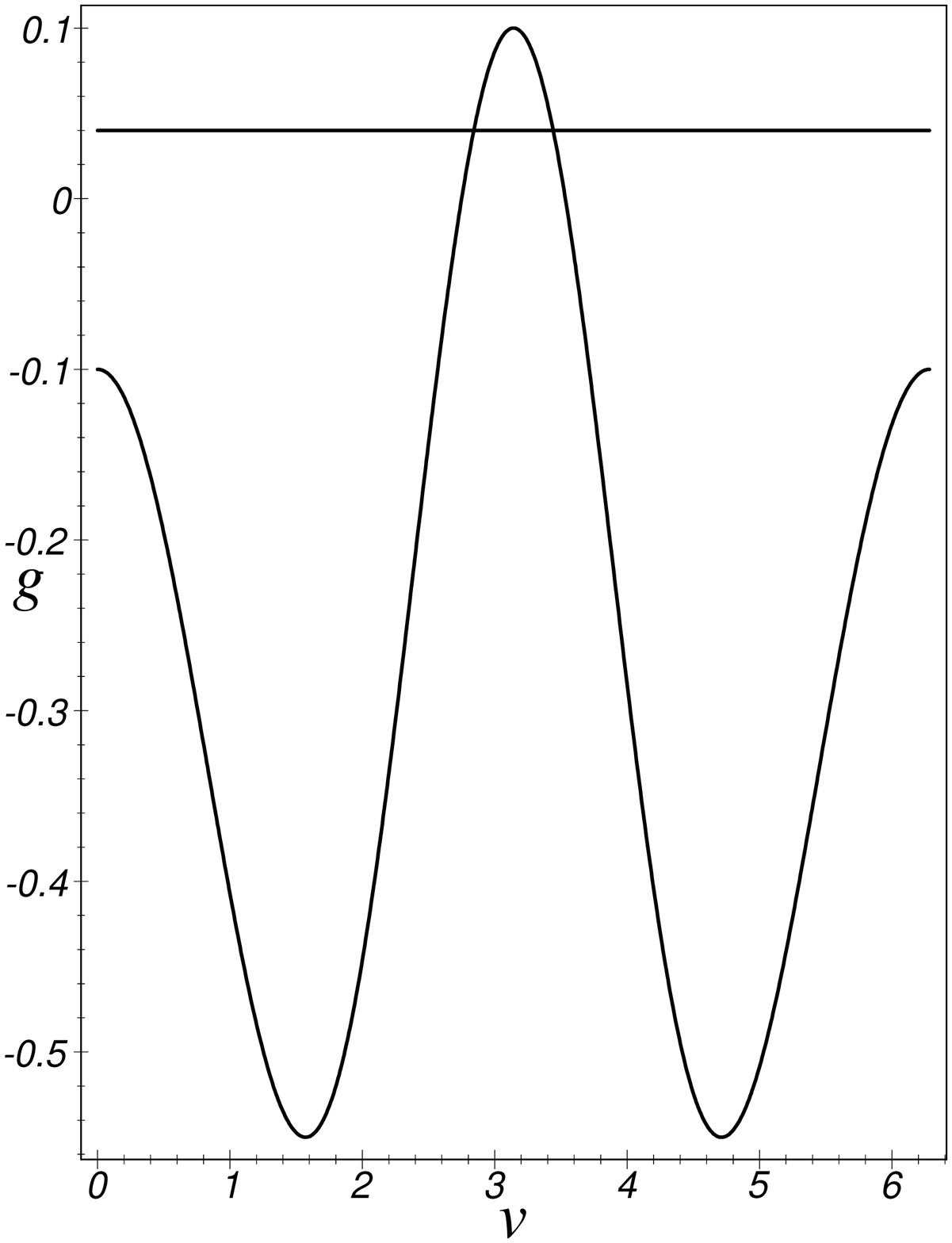}\hfill
            \epsfxsize=1.7in\epsfbox{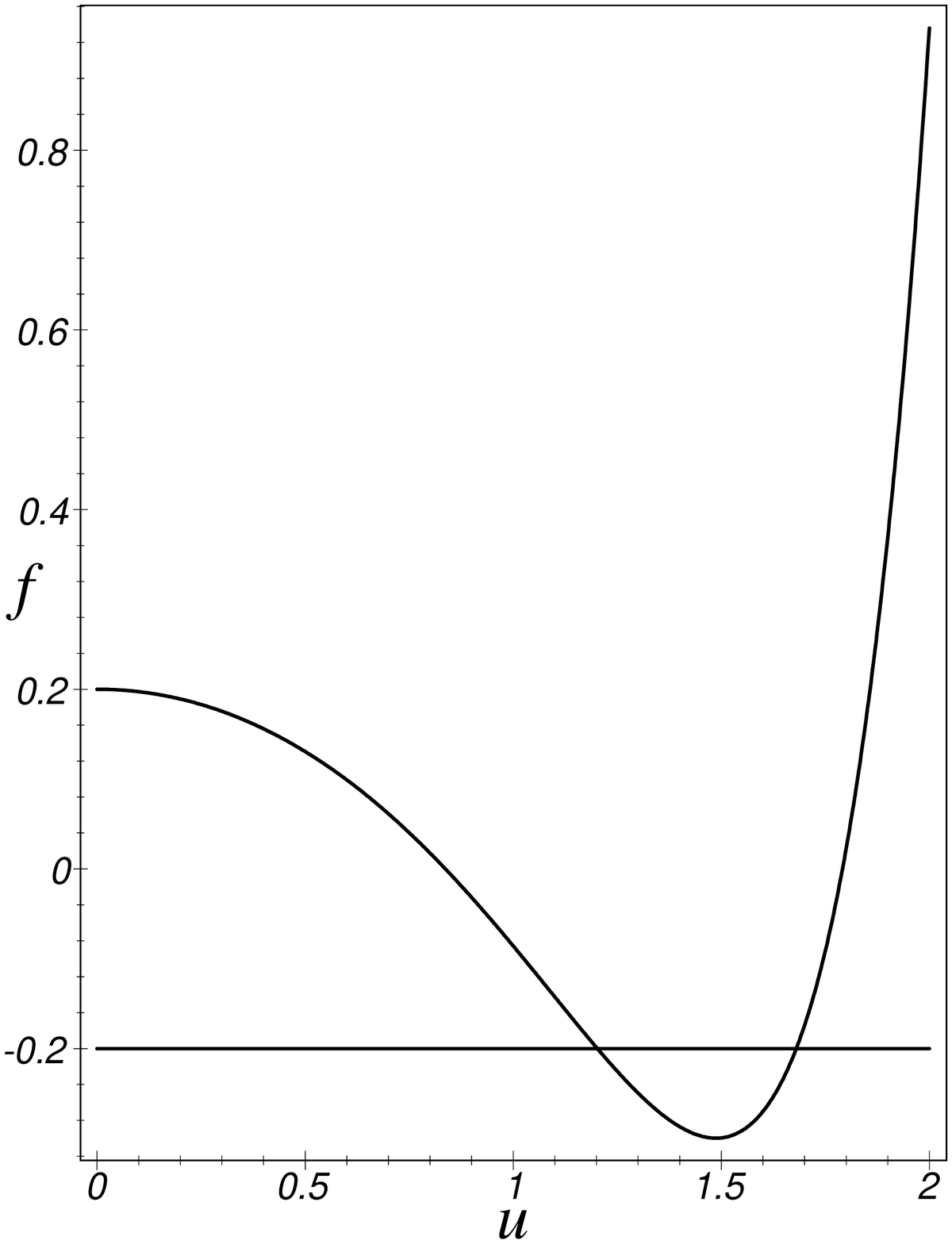}
            \epsfxsize=1.7in\epsfbox{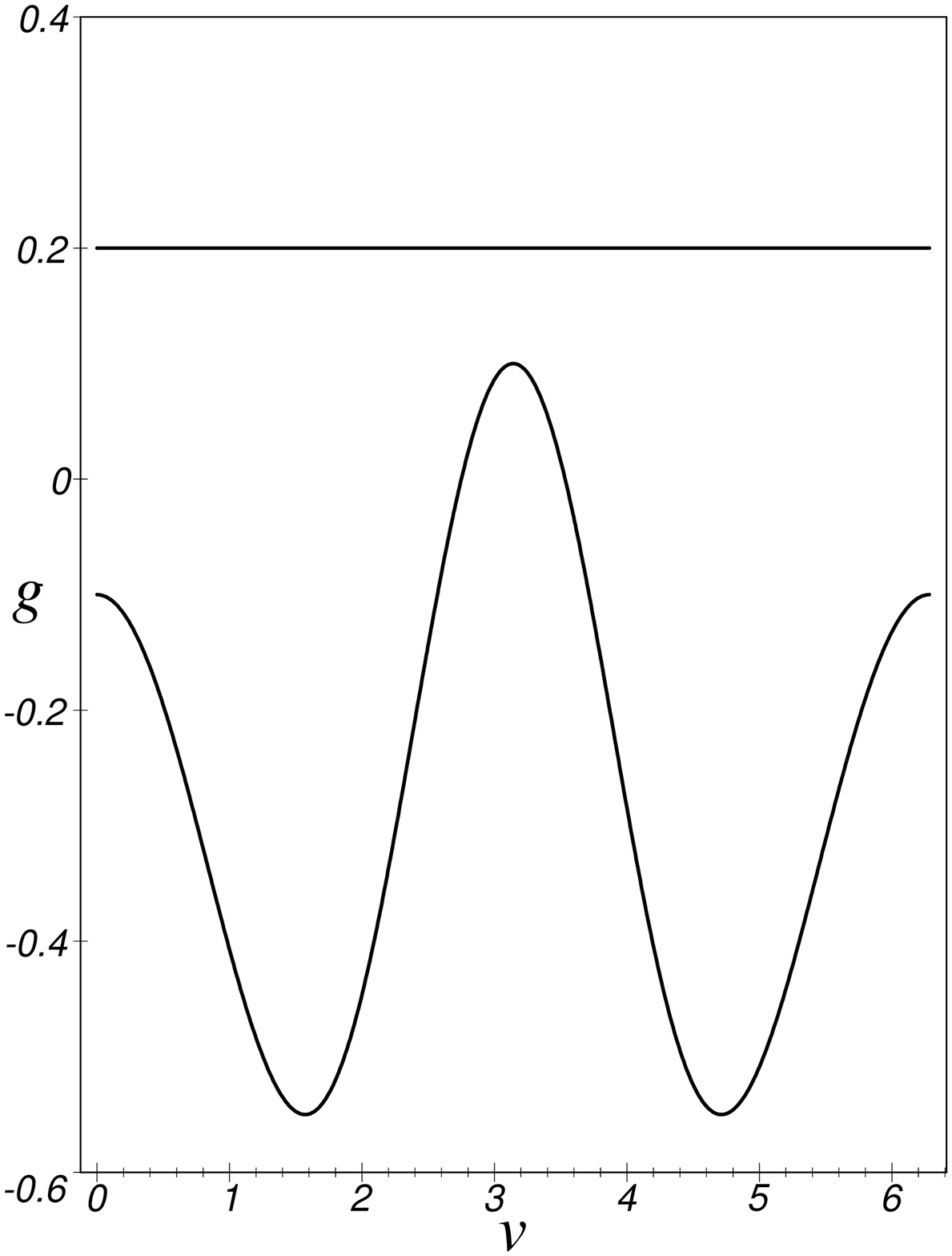}}
\centerline{\hspace*{0.8in}$(e)$\hspace*{1.5in}$(f)$\hfill
            $(g)$\hspace*{1.5in}$(h)$\hspace*{0.8in}}
\caption[Figure 3]{The graphs of $f(u)$ and $g(v)$ for $\beta=0.5$,
$\gamma=2.8$, $C=0.2$, $E=1.1$ and $a=0.5$. The horizontal
lines indicate the levels of $I_2$ and $-I_2$ in the graphs
of $f(u)$ and $g(v)$, respectively.
(a) $I_2=0.3$ (b) $-I_2=-0.3$ (c) $I_2=0.15$
(d) $-I_2=-0.15$ (e) $I_2=-0.04$ (f) $-I_2=0.04$ (g) $I_2=-0.2$ 
(h) $-I_2=0.2$.} 
\end{figure*}

\section{ORBIT FAMILIES}
We carry out a transformation to elliptic coordinates
as follows
\begin{equation}
x=a \cosh u \cos v, \label{13}
\end{equation}
\begin{equation}
y=a \sinh u \sin v, \label{14}
\end{equation}
\noi where $u \geq 0$ and $0\le v \le 2 \pi$.
The curves of constant $u$ and $v$ are confocal
ellipses and hyperbolas, respectively. P1 and P2
are the foci of these curves. In the new coordinates, 
the motion of a test star is determined by the Hamiltonian
\begin{equation}
{\cal H} = \frac 1{2a^2(\sinh ^2 u + \sin ^2 v)} 
(p_u^2+p_v^2) + \Phi (u,v), \label{15}
\end{equation}
\noi where $p_u$ and $p_v$ denote the canonical momenta and 
\begin{eqnarray}
\Phi &=& \frac {F(u)+G(v)}{2a^2(\sinh ^2 u + \sin ^2 v)},
\label{16} \\
F(u) &=& C (\cosh u)^{\gamma}, \label{17} \\
G(v) &=& -C \beta \cos v |\cos v|^{\gamma-1}, \label{18} \\
C &=& K (2a)^{\gamma}. \nonumber
\end{eqnarray}
\noi The transformed potential (\ref{16})
is of St\"ackel form for which the Hamilton-Jacobi
equation separates and yields the second integral of
motion, $I_2$. We obtain
\begin{equation}
I_2=p_u^2 - 2a^2 E \sinh ^2 u + F(u),  \label{19}
\end{equation}
\noi or equivalently
\begin{equation}
-I_2=p_v^2 - 2a^2 E \sin ^2 v + G(v),  \label{20}
\end{equation}
\noi where $E$ is the total energy of the system,
$E \equiv {\cal H}$. The potential function ($\Phi$) 
is positive everywhere. Hence, we immediately 
conclude $E>0$. 

Having the two isolating integrals $E$ and $I_2$, one can find
the possible regions of motion by employing the positiveness
of $p_u^2$ and $p_v^2$ in (\ref{19}) and (\ref{20}).
We define the following functions:
\begin{eqnarray}
f(u)&=& -2 a^2 E \sinh ^2 u + F(u), \label{21} \\
g(v)&=& -2 a^2 E \sin ^2 v + G(v). \label{22}
\end{eqnarray}
\noi Since $p_u^2 \geq 0$ and $p_v^2 \geq 0$, one can write
\begin{eqnarray}
I_2-f(u) &\geq& 0, \label{23} \\
-I_2-g(v) &\geq& 0. \label{24}
\end{eqnarray}

\noi Orbits are classified based on the behaviour of 
$f(u)$ and $g(v)$. The most general form of $f(u)$ 
is attained for $\gamma C<4a^2E$. In such a circumstance, 
$f(u)$ has a local maximum at $u=0$, $f_{\rm M}=f(0)=C$, 
and a global minimum at $u=u_{\rm m}$, $f_{\rm m}=f(u_{\rm m})$,  
where 
\begin{equation}
\cosh u_{\rm m}=\left ( \frac {4a^2E}{C\gamma} \right )^
{\frac {1}{\gamma-2}}, \label{25}
\end{equation}
\noi and
\begin{equation}
f_{\rm m}=-2a^2E \sinh ^2 u_{\rm m} + 
C (\cosh u_{\rm m})^{\gamma}. \label{26} 
\end{equation}
\noi From (\ref{23}) we obtain 
\begin{equation}
I_2 \ge f_{\rm m}. \label{27}
\end{equation}
On the other hand, $g(v)$ has a global maximum at 
$v=\pi$, $g_{\rm M}=g(\pi)=\beta C$, and two global 
minima at $v=\pi/2$ and $v=3\pi/2$, 
$g_{\rm m}$=$g(\pi/2)$=$g(3\pi/2)$=$-2a^2E$.
Therefore, Inequality (\ref{24}) implies
\begin{equation}
I_2 \le 2a^2E. \label{28} 
\end{equation}
By combining (\ref{27}) and (\ref{28}) one achieves
\begin{equation}
f_{\rm m} \le I_2 \le 2a^2E. \label{29} 
\end{equation}
\noi By taking $2< \gamma < 3$ and $\gamma C<4a^2E$ 
into account, we arrive at $2a^2E>C$. Furthermore, 
$f_{\rm m}$ and in consequence $I_2$, can take both 
positive and negative values. For a specified value 
of $E$, the following types of orbits occur as $I_2$ 
varies.

(i) {\it Butterflies}. For $C \le I_2 <2a^2E$, 
the allowed values for $u$ and $v$ are
\begin{equation}
u \le u_0,~v_{b,1} \le v \le v_{b,2},
~v_{b,3} \le v \le v_{b,4}, \label{30}
\end{equation}
\noi where $u_0$ and $v_{b,i}$ ($i=1,2,3,4$) are the
roots of $f(u)=I_2$ and $g(v)=-I_2$, respectively.
As Fig.~3a shows, the horizontal line that indicates
the level of $I_2$, intersects the graph of $f(u)$ at
one point, which specifies the value of $u_0$. The
line corresponding to the level of $-I_2$ intersects
$g(v)$ at four points that give the values of
$v_{b,i}$s (Fig.~3b). In this case the motion
takes place in a region bounded by the coordinate
curves $u=u_0$ and $v=v_{b,i}$. The orbits fill
the shaded region of Fig.~4a. These are butterfly
orbits (de Zeeuw 1985) that appear around the
local minimum of $\Phi$ at ($x=0,y=0$).

(ii) {\it Nucleuphilic Bananas}. For $\beta C \le I_2 < C$
the equation $f(u)=I_2$ has two roots, $u_{n,1}$ and $u_{n,2}$, 
which can be identified by the intersections of $f(u)$ 
and the level line of $I_2$ (see Fig.~3c). In this case,
the equation $g(v)=-I_2$ has four real roots, 
$v=v_{n,i}$ ($i=1,2,3,4$), (Fig.~3d). 
The allowed ranges of $u$ and $v$ will be
\begin{equation}
u_{n,1} \le u \le u_{n,2},~
v_{n,1} \le v \le v_{n,2},~
v_{n,3} \le v \le v_{n,4}. \label{31}
\end{equation}
\noi The orbits (Fig.~4b) are bound to the curves of 
$u=u_{n,1}$, $u=u_{n,2}$ and $v=v_{n,i}$. We call them 
nucleuphilic banana orbits, for they look like banana 
and bend toward the nuclei.  

\begin{figure}
\centerline{\hbox{\epsfxsize=1.7in\epsfbox{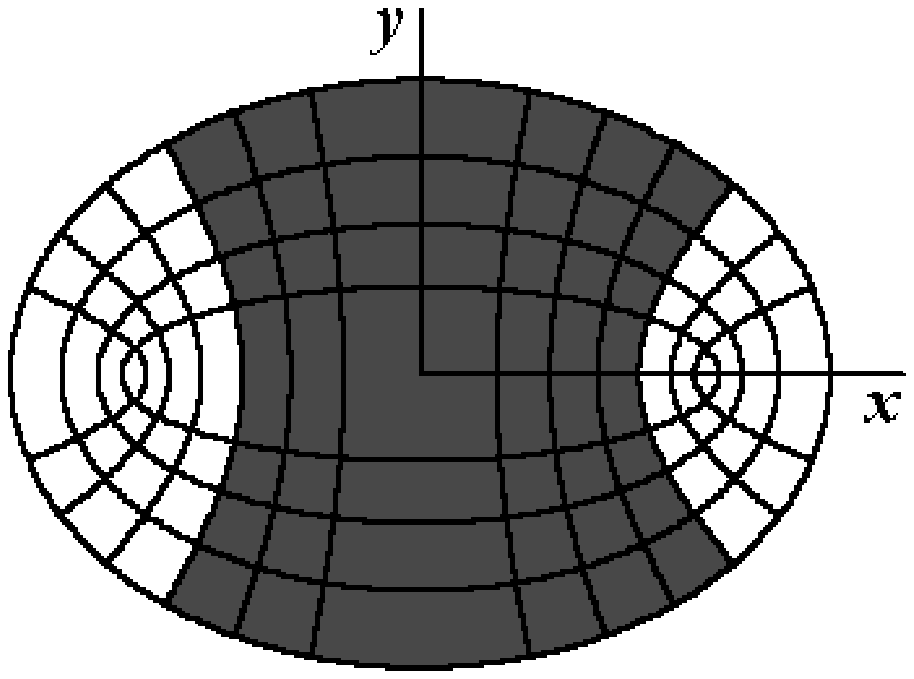}
                  \epsfxsize=1.7in\epsfbox{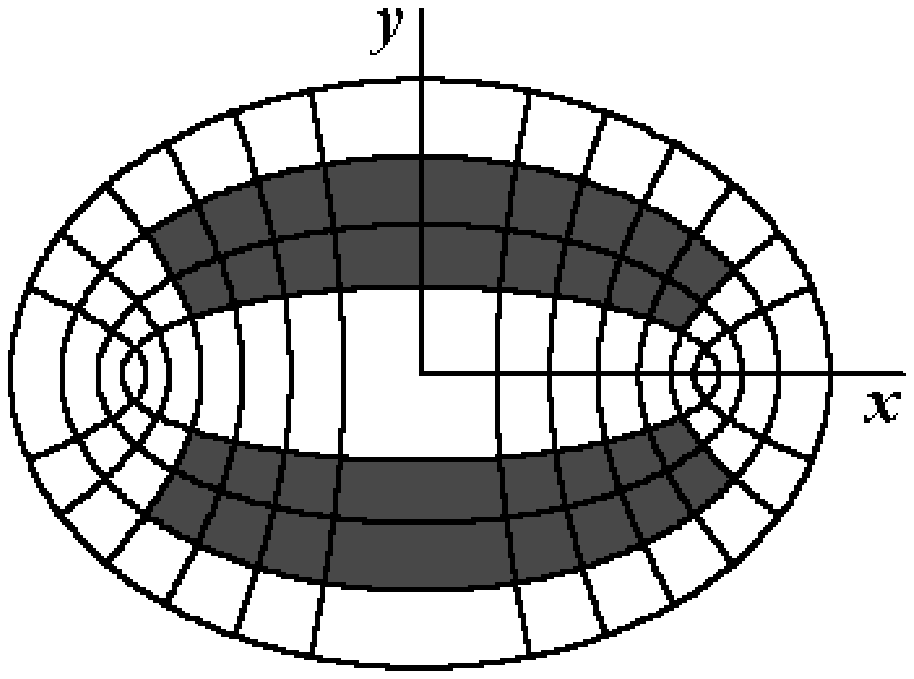}}}
\centerline{\hspace*{0.75in}$(a)$\hfill$(b)$\hspace{0.75in}}
\vspace{0.3cm}
\centerline{\hbox{\epsfxsize=1.7in\epsfbox{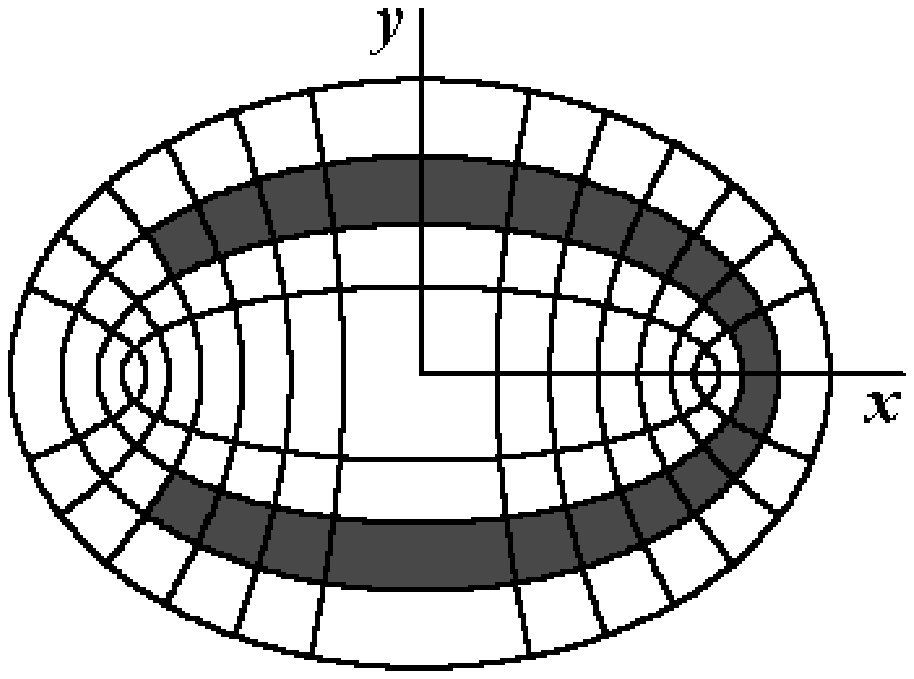}
                  \epsfxsize=1.7in\epsfbox{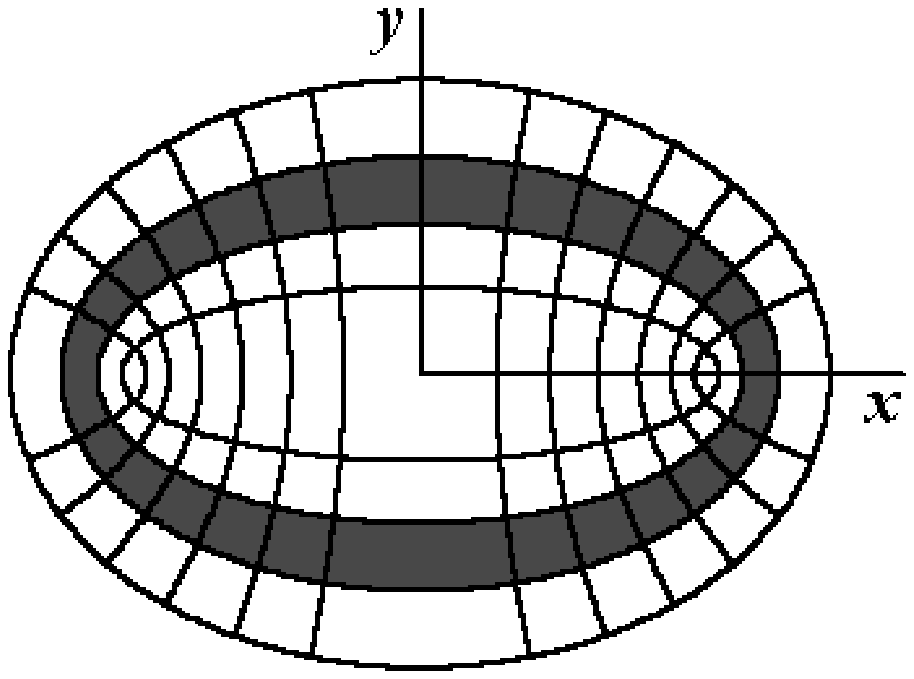}}}
\centerline{\hspace*{0.75in}$(c)$\hfill$(d)$\hspace{0.75in}}
\vspace{0.3cm}
\centerline{\hbox{\epsfxsize=1.7in\epsfbox{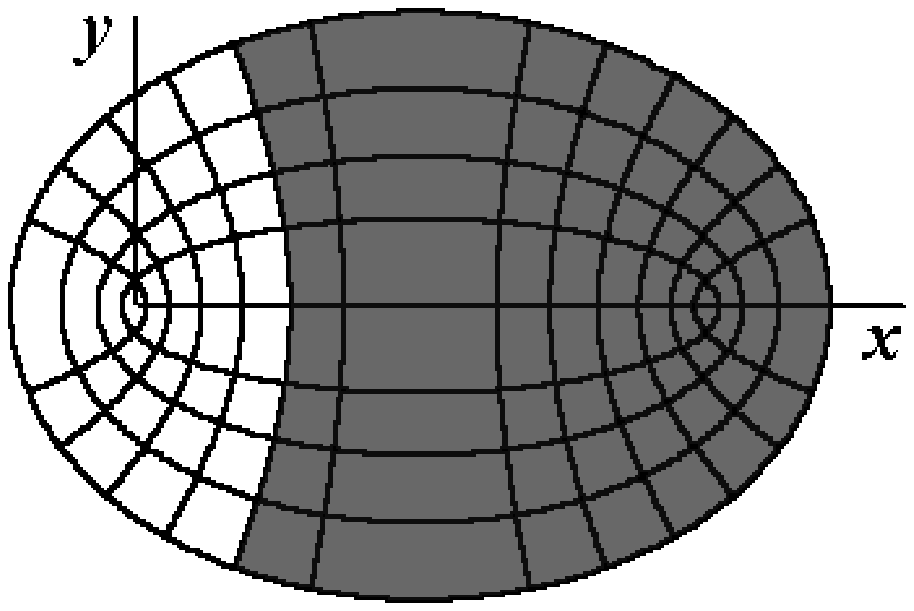}}}
\centerline{$(e)$}
\caption[Figure 4]{The possible families of orbits:
(a) a butterfly orbit (b) nucleuphilic banana orbits
(c) a horseshoe orbit (d) an aligned loop orbit (e)
a lens orbit associated with $\beta=1$ and $I_2=C$.
For $\beta \not = 0$, the orbits are only symmetric 
with respect to the $x$-axis. Loop orbits are exceptional;
they are always symmetric with respect to the coordinate axes.} 
\end{figure}

(iii) {\it Horseshoes}. For $-\beta C \le I_2<\beta C$, 
both of the equations $f(u)=I_2$ and $g(v)=-I_2$ have 
two real roots. We denote these roots by $u=u_{h,i}$ 
and $v=v_{h,i}$ ($i=1,2$). 
In other words, the level lines of $\pm I_2$ 
intersect the graphs of $f(u)$ and $g(v)$ at two points 
as shown in Figs.~3e and 3f. The trajectories of 
stars fill the shaded region of Fig.~4c. 
We call these horseshoe orbits.

(iv) {\it Aligned Loops}. For $f_m < I_2<-\beta C$, 
the equation $f(u)=I_2$ has two real roots, $u=u_{l,i}$ 
($i=1,2$) while the equation $g(v)=-I_2$ has
no real roots and Inequality (\ref{24}) is always 
satisfied (Figs.~3g and 3h). The orbits fill a tubular 
region as shown in Fig.~4d. We call these aligned loops 
because they are aligned with the surface density 
isocontours of outer regions. 

(v) {\it Transitional cases}. For $I_2=2a^2E$, stars
undergo a rectilinear motion on the $y$-axis with the
amplitude of $\pm a \sinh u_0$. For $I_2=f_{\rm m}$,
loop orbits are squeezed to an elliptical orbit defined
by $u=u_{\rm m}$. For $\beta=0$, horseshoe orbits are
absent, leaving the other types of orbits symmetric
with respect to the coordinate axes. Banana orbits no
longer survive for $\beta=1$ (eccentric disc model).
In this case, butterflies extend to a {\it lens} orbit
when $I_2=C$ (see Figure 4e).
For $\gamma C > 4a^2E$, $f(u)$ is a monotonically
increasing function of $u$ and ``low-energy"
butterflies are the only existing family of orbits. 
These are small-amplitude liberations in the vicinity 
of the local minimum of $\Phi$ at $(x=0,y=0)$.

\section{THE POSITIVENESS OF THE SURFACE DENSITY}
The sign of $\Sigma$ is linked to that of $\nabla ^2 \Phi$ through
Equation (\ref{3}). To prove that $\Sigma$ takes positive values for
the potentials of (\ref{1}), it suffices to show that the Laplacian 
of $\Phi$ is a positive function of $\gamma$, $\beta$, $u$ and $v$.

Consider the Laplace equation in elliptic coordinates as
\begin{eqnarray}
\nabla ^2 \Phi &=& \frac {1}{a^2D}
\left ( \Phi _{,uu} + \Phi _{,vv} \right ), \label{laplace} \\
D &=& \sinh ^2u + \sin ^2v, \nonumber
\end{eqnarray}
where $_{,s}$ denotes $\frac {\partial}{\partial s}$.
Substituting from (\ref{16}) into (\ref{laplace}), leads to 
\begin{equation}
\nabla ^2 \Phi = \frac {{\cal F}(\gamma,\beta;u,v)} {2a^4D^4}, \label{feq}
\end{equation}
with
\begin{eqnarray}
{\cal F} &=& D^2(F_{,uu}+G_{,vv})-D(F+G)(D_{,uu}+D_{,vv})
\nonumber \\
&{}& -2D(F_{,u}D_{,u}+G_{,v}D_{,v})\nonumber \\
&{}& +2(F+G)(D_{,u}^2+D_{,v}^2). \label{F_eq}
\end{eqnarray}
For the sake of simplicity, we assume $C=1$.
We show that the minimum of ${\cal F}$ is always positive.
We prove our claim for 
$-\frac {\pi}{2}\leq v \leq \frac {\pi}{2}$, which implies 
$G(v)=-\beta \cos ^{\gamma}v$ (a similar method can be repeated
for $\frac {\pi}{2}< v < \frac {3\pi}{2}$). In this case,
${\cal F}$ will be a linear, decreasing function of $\beta$
(because $\Phi$ has such a property).
Therefore, one concludes 
${\cal F}(\gamma,1;u,v) \leq {\cal F}(\gamma,\beta;u,v)$.
Furthermore, ${\cal F}$ directly depends on $\cosh u$,
which results in ${\cal F}(\gamma,1;0,v) \leq {\cal F}(\gamma,1;u,v)$.
Hence, $\nabla ^2 \Phi$ is positive if 
${\cal G}(\gamma,v)\equiv {\cal F}(\gamma,1;0,v) \geq 0$.
By the evaluation of (\ref{F_eq}) for $\beta=1$ and $u=0$, 
one finds out
\begin{eqnarray}
{\cal G}(\gamma,v) &=& -\gamma ^2 \sin ^6 v \cos ^{\gamma -1}v +
\sin ^2 v (1-\cos ^{\gamma}v) \nonumber \\ 
&{}& + \gamma [\sin ^4 v + \cos ^{\gamma-2}v(\sin ^6 v \nonumber \\
&{}& - 3 \cos ^2 v \sin ^4 v)]. \label{G_gam_v}
\end{eqnarray}
We have plotted ${\cal G}(\gamma,v)$ in Figure 5. On the evidence
of this figure, ${\cal G}$ is a positive function for
$-\frac {\pi}{2}\leq v \leq \frac{\pi}{2}$ and $2<\gamma <3$.
Thus, the surface density distribution takes positive values 
for all of our model galaxies.

\begin{figure}
\centerline{\hbox{\epsfxsize=1.7in\epsfbox{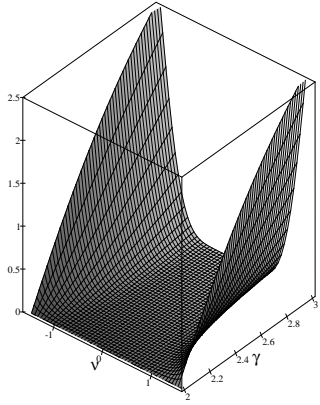}}}
\caption[Figure 5]{The behaviour of ${\cal G}(\gamma,v)$ 
for $-\frac {\pi}2 \leq v \leq \frac {\pi}2$ and $2<\gamma <3$.} 
\end{figure}

\section{DISCUSSION}
In his pioneering work, Euler showed 
the separability of motion in the potential field of 
two {\it fixed} Newtonian centres of attraction. This problem
was then completely solved by Jacobi (Pars 1965). It 
is physically impossible to keep apart these two ``point 
masses", for they will attract each other leading to
an eventual collapse. However, the assumed masses 
can be in equilibrium if they revolve around their
common centre of mass (this is the classical 
3-body problem). Our planar model is indeed 
Jacobi's problem in which we have replaced 
two fixed centres of gravitation with a continuous
distribution of matter, where mass concentration increases 
towards two nuclei (P1 and P2) in power-law, strong cusps. 
These nuclei are maintained by an interesting family of 
orbits, nucleuphilic bananas. Below, we explain why the 
mentioned nuclei are generated and don't collapse. 

The force exerted on a star is equal to $-\nabla \Phi$. 
The motion under the influence of this force can be 
tracked on the {\it potential hill} of Figure 2a. 
This helps us to better imagine the motion trajectories.

\subsection{The behaviour of orbits}
Stars moving in nucleuphilic banana orbits undergo 
motions {\it near} the 1:2 resonance. They oscillate 
twice in the $y$-direction for each $x$-axial 
oscillation. The turning points of this group of 
stars lie on the curves $v=v_{n,i}$.
These hyperbolic curves can be chosen arbitrarily
close to P1 or P2. When stars approach P1 (or P2),
their motion slows down (because they climb on the 
cuspy portion of the potential hill and considerably 
lose their kinetic energy while the potential
energy takes a maximum) and the orbital angular
momentum switches sign somewhere on $v=v_{n,i}$.
Thus, these stars spend much time in the vicinity
of P1 (or P2) and deposit a large amount of mass.
This phenomenon is the main reason for the
generation of cuspy zones around P1 and P2.
Stars moving in nucleuphilic bananas cross the 
$y$-axis quickly, and therefore, don't 
contribute much mass to the region between the 
nuclei.

Horseshoe orbits cause the sizes of cuspy zones 
to be different through the following mechanism.
Stars that start their motion sufficiently close
to P1 (larger nucleus), are repelled from
P1 because the force vector is not directed
inward in this region. As they move outward,
their orbits are bent and cross the $x$-axis
with non-zero angular momentum. These stars
linger only near P1, and in consequence, help
the cuspy zone around P1 grow more than that
of P2. The asymmetry of nucleuphilic bananas,
with respect to the $y$-axis, is also an origin
of the different sizes of cuspy zones. In fact,
horseshoe orbits are born once nucleuphilic
bananas join together for $I_2=\beta C$.
Horseshoe and nucleuphilic banana orbits are
the especial classes of boxlets that appropriately
bend toward the nuclei. The lack of such a property
in centrophobic banana orbits causes the discs of
Sirdhar \& Touma (1997) to be non-self-consistent.

Aligned loop orbits occur when the orbital angular
momentum is high enough to prevent the test particle
to slip down on the potential hill. The boundaries
of loop orbits are defined by the ellipses
$u=u_{l,1}$ and $u=u_{l,2}$. The nuclear cusps are
located at the foci of these ellipses.
Aligned loops have the same orientation as the
surface density isocontours (compare Figures 1 and
4d). Thus, according to the results of Z99, it is 
possible to construct a self-consistent model using 
aligned loop orbits.

It is worthy to note that butterfly orbits play a 
significant role in maintaining the non-axisymmetric
structure of the model at the moderate distances
of ${\cal O}(a)$.

\subsection{The nature of P1 and P2}
The points where the cusps have been located,
are inherently unstable. With a small
disturbance, stars located at ($x=\pm a,y=0$) are
repelled from these points because $-\nabla \Phi$
is directed outward when $r_i \rightarrow 0$
($i=1,2$). But, the time that stars spend near the
nuclei will be much longer than that of distant
regions when they move in horseshoe and banana orbits.
The points P1 and P2 are unreachable, for they 
correspond to the energy level $E=+\infty$. Based on 
the results of this paper, we conjecture that there 
may not be any mass concentration just at the centre of 
cuspy galaxies. However, a very dense region exists 
{\it arbitrarily} close to the centre!

\subsection{The double nucleus can be in equilibrium}
The nuclei pull each other due to their mutual
gravitational attraction and it seems that they must
collapse. However, we explain that in certain circumstances,
the double nucleus can be in {\it static} equilibrium.
At first we estimate the mass inside the separatrices of
the surface density distribution (the mass of cuspy zones)
and concentrate the matter at P1 and P2 (this
is logical because the surface density distribution is
almost axisymmetric near the nuclei).
In this way, we obtain two point masses, $M_1$ and $M_2$.
According to (\ref{7}) and (\ref{9}), the following relations
approximately hold 
\begin{eqnarray}
M_i &=& \int \nolimits _{-\pi}^{\pi}
        \int \nolimits _{\epsilon}^{r_{0i}}
        \sigma _{i} r^{-1} {\rm d}r {\rm d}\theta, 
~\epsilon \rightarrow 0, \nonumber \\
    &=& 2 \pi \sigma _i \log {\frac {r_{0i}}{\epsilon}},
        ~~i=1,2, \label{32}
\end{eqnarray}
\noi where $r_{0i}$ are chosen as the radii of inner tangent
circles to the separatrices and the constant parameters 
$\sigma _i$ are computed based on the surface density profile 
near the nuclei. As $\epsilon \rightarrow 0$, $M_i$s diverge 
to infinity unless a negative mechanism prevents them to 
grow. Consider the discs of radius $\epsilon$ with the 
centres located at $(\pm a,0)$ and call them ${\cal D}_1$ 
and ${\cal D}_2$. Since P1 and P2 are ``locally" unstable, 
we rely on our previous argument that the matter is swept 
out from these points, allowing us to exclude ${\cal D}_1$ 
and ${\cal D}_2$ from our model for some $0< \epsilon \ll 1$. 
In this way, $M_i$s take finite values. 

\begin{figure}
\centerline{\hbox{\epsfxsize=1.7in\epsfbox{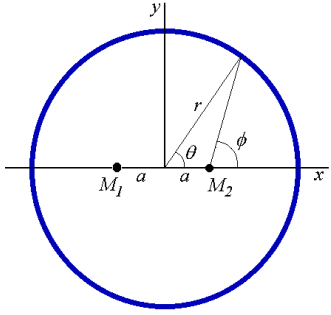}}}
\caption[Figure 5]{In this figure we have shown a circular
ring of matter of outer regions. The masses of cuspy zones
can approximately be computed and concentrated at the points
($\pm a,0$). The gravitational attraction of the ring upon
the point mass $M_1$ ($M_2$) is directed in the $x$-direction
due to the existing symmetry.} 
\end{figure}

$M_1$ and $M_2$ attract each other and start to move 
if they are not influenced by other gravitational sources.
We claim that the required extra force comes from the
gravitational attraction of the matter of outer regions.
Consider Figure 6 where $M_1$ and $M_2$ are shown along
with a ring of matter of outer regions. For brevity,
we assume $\beta=0$, which yields $M_1=M_2=M$. Due to the 
existing symmetry, the gravitational force exerted on 
$M_2$ by the assumed ring will have a resultant in the 
$x$-direction. When $r$ is sufficiently large, $r \gg a$, 
this force is calculated as follows
\begin{eqnarray}
F_x(r) &=& GM \sigma _{\infty} \int\nolimits_{-\pi}^{\pi}
\frac {r^{\gamma -2} \cos \phi {\rm d} \theta}
{r^2+a^2-2ar\cos \theta}, \label{33} \\
\cos \phi &=& \frac {r\cos \theta -a}
{(r^2+a^2-2ar\cos \theta)^{1/2}}, \label{34}
\end{eqnarray}
where we have used 
$\Sigma \approx \sigma _{\infty} r^{\gamma -3}$ with
$\sigma _{\infty}$ being a positive constant (see Eq. (\ref{11})).
By integrating $F_x(r)$ over $r$ from 
some $r=R \gg a$ to $r=\infty$, the total force, due 
to the matter of outer regions, is found to be
\begin{equation}
F_x = GM \sigma _{\infty} \int\nolimits_{R}^{\infty} 
\int\nolimits_{-\pi}^{\pi}
\frac {r^{\gamma -2} (r\cos \theta -a) {\rm d} \theta {\rm d}r}
{(r^2+a^2-2ar\cos \theta)^{3/2}}. \label{35}
\end{equation}
By a change of independent variable as $\xi =a/r$, the integrand
can be simplified in the form 
\begin{eqnarray}
F_x &=& b \int\nolimits_{0}^{\xi _0} \xi ^{2-\gamma} {\rm d}\xi
          \frac {{\rm d}}{{\rm d}\xi} \int\nolimits_{0}^{\pi}
          \frac {{\rm d}\theta}{(1+\xi ^2-2\xi \cos \theta)^{1/2}},
          \label{36} \\
b  &=& \frac {2GM \sigma _{\infty}}{a^{3-\gamma}}, \label{37}
\end{eqnarray}
where $\xi_0=a/R$. Consequently,
\begin{equation}
F_x = b \int\nolimits_{0}^{\xi _0} \xi ^{2-\gamma} {\rm d}\xi
      \frac {{\rm d}}{{\rm d}\xi} \sum _{n=0}^{\infty} \xi ^n
      \int\nolimits_{0}^{\pi} P_n(\cos \theta) {\rm d}\theta,
      \label{38} 
\end{equation}
with $P_n$s being the well known Legendre functions.
According to (Morse \& Feshbach 1953)
\begin{eqnarray}
\int\nolimits_{0}^{\pi} P_{2k+1}(\cos \theta)
{\rm d} \theta &=& 0, \label{39} \\
\int\nolimits_{0}^{\pi} P_{2k}(\cos \theta)
{\rm d} \theta &=& \pi \left [ \frac {(2k)!}{(2^k k!)^2} 
\right ] ^2 \equiv c_{2k}, \label{40}
\end{eqnarray}
one achieves
\begin{equation}
F_x = b \int\nolimits_{0}^{\xi _0} (\xi ^{2-\gamma} 
        \sum _{k=1}^{\infty} 2k c_{2k} \xi ^{2k-1})
        {\rm d}\xi. \label{41} 
\end{equation}
Integrating (\ref{41}) over $\xi$, yields
\begin{eqnarray}
F_x &=& b Q(\xi _0), \label{42} \\
Q(\xi _0) &=& \xi _0 ^{2-\gamma}
              \sum _{k=1}^{\infty}
              \frac {2k c_{2k}}{2k+2-\gamma}
              \xi _0 ^{2k}. \label{43}
\end{eqnarray}
It is obvious that $Q$ is a positive function of $\xi _0$.
Therefore, from (\ref{42}) one concludes $F_x>0$ indicating 
that $M_2$ is pulled away from the centre. 
$M_2$ will be in static equilibrium if $F_x$ is
balanced with the gravitational force of $M_1$, i.e.,
\begin{equation}
F_x=\frac {GM_1M_2}{(2a)^2}=\frac {GM^2}{(2a)^2}. \label{44}
\end{equation}
By substituting from (\ref{32}) and (\ref{42}) into (\ref{44}),
we obtain
\begin{equation}
\delta \equiv \frac {\epsilon}{r_{01}} = e^{-s},~~s>0, \label{45}
\end{equation}
where $r_{01}=r_{02}$, $\sigma _1=\sigma _2$ 
(because we assumed $\beta=0$) and
\begin{equation}
s=\frac {4 \sigma _{\infty} a^{\gamma -1} Q(\xi _0)}
  {\pi \sigma _1}. \label{46}
\end{equation}
Our numerical computations of $\Sigma$ reveal that 
$\sigma_{\infty} \gg \sigma_i$, which guarantees 
$\epsilon \ll r_{01}$ as desired. Following a similar
procedure as above, one can show that the double 
nucleus remains in equilibrium for $\beta >0$. 

\subsection{The nuclei of M31 and NGC4486B}
In many respects, the surface density isocontours of
our model galaxies are similar to the isophotal lines
of the nuclei of M31 and NGC4486B. Our mass models
are cuspy within two separatrices. Such curves can 
be distinguished in the nuclei of M31 
and NGC4486B (see L96 and T95).
We are not sure that the nuclei of M31 and NGC4486B are 
really cuspy, because existing telescopes can not 
highly resolve the regions around P1 and P2 (even HST 
images contain ``few" bright pixels at the locations of 
P1 and P2). Whatever the mass distribution inside these 
nuclei may be, our models reveal that double nuclei can 
exist even in the absence of supermassive BHs. 

The nucleus of M31 can also be explained by the
eccentric disc corresponding to $\beta=1$.
In such a circumstance, stars moving in butterfly orbits
form a local group in the vicinity of ($x=0,y=0$).
The accumulation of stars around this local minimum
of $\Phi$ can create a faint nucleus like P2 (see T95).
Therefore, P2 will approximately be located at the ``centre"
of loop orbits while the eccentric, brighter nucleus (P1)
is at the location of the cusp. In other words, loop and
high-energy butterfly orbits will control the overall shape
of outer regions, horseshoe orbits will generate P1 and
low-energy butterflies will create P2.

\subsection{Challenging problems}
It is not known for us if there are rotationally
supported double structures or not. This idea comes
from the fact that we can replace the point masses of
the restricted 3-body problem (the restricted 3-body 
problem is usually expressed in a rotating frame) 
with a continuous distribution of matter. Moreover, 
NGC4486B and the bulge of M31, are three dimensional 
objects and the assumption of planar models seems
to be a severe constraint.

So far we showed that the double nucleus can be in 
static equilibrium due to the existing gravitational 
effects of the model. The stability study of such
states, however, remains as a challenging problem.

Our next goal is to apply the method of Schwarzschild
(1979,1993) for the investigation of self-consistency.

\section{ACKNOWLEDGMENTS}
The authors wish to thank the anonymous referee for
illuminating questions and valuable comments on the
paper.

\end{document}